\documentclass[sigconf, balance=false]{acmart}
\usepackage{popets}
\usepackage{multirow}
\usepackage{dblfloatfix}
\usepackage{float}
\usepackage{wrapfig}
\usepackage{listings}
\usepackage[inline,shortlabels]{enumitem}
\usepackage{color}
\usepackage{xspace}
\usepackage{subcaption}
\usepackage{framed}
\usepackage{algorithmic}

\lstdefinelanguage{JavaScript}{
  morekeywords=[1]{break, continue, delete, else, for, function, if, in,
    new, return, this, typeof, var, void, while, with},
  morekeywords=[2]{false, null, true, boolean, number, undefined,
    Array, Boolean, Date, Math, Number, String, Object},
  morekeywords=[3]{eval, parseInt, parseFloat, escape, unescape},
  sensitive,
  morecomment=[s]{/*}{*/},
  morecomment=[l]//,
  morecomment=[s]{/**}{*/},
  morestring=[b]',
  morestring=[b]"
}[keywords, comments, strings]
\lstalias[]{ES6}[ECMAScript2015]{JavaScript}
\lstdefinelanguage[ECMAScript2015]{JavaScript}[]{JavaScript}{
  morekeywords=[1]{await, async, case, catch, class, const, default, do,
    enum, export, extends, finally, from, implements, import, instanceof,
    let, static, super, switch, throw, try},
  morestring=[b]` 
}
\definecolor{mediumgray}{rgb}{0.3, 0.4, 0.4}
\definecolor{mediumblue}{rgb}{0.0, 0.0, 0.8}
\definecolor{forestgreen}{rgb}{0.13, 0.55, 0.13}
\definecolor{darkviolet}{rgb}{0.58, 0.0, 0.83}
\definecolor{royalblue}{rgb}{0.25, 0.41, 0.88}
\definecolor{crimson}{rgb}{0.86, 0.8, 0.24}

\lstdefinestyle{JSES6Base}{
  backgroundcolor=\color{white},
  basicstyle=\footnotesize\ttfamily,
  breakatwhitespace=false,
  breaklines=false,
  captionpos=b,
  columns=fullflexible,
  commentstyle=\color{mediumgray}\upshape,
  emph={},
  emphstyle=\color{crimson},
  extendedchars=true,  % requires inputenc
  fontadjust=true,
  frame=none,
  identifierstyle=\color{black},
  keepspaces=true,
  keywordstyle=\color{mediumblue},
  keywordstyle={[2]\color{darkviolet}},
  keywordstyle={[3]\color{royalblue}},
  numbers=none,
  numbersep=5pt,
  numberstyle=\tiny\color{black},
  rulecolor=\color{black},
  showlines=true,
  showspaces=false,
  showstringspaces=false,
  showtabs=false,
  stringstyle=\color{forestgreen},
  tabsize=2,
  title=\lstname,
  upquote=true  % requires textcomp
}

\lstdefinestyle{JavaScript}{
  language=JavaScript,
  style=JSES6Base
}
\lstdefinestyle{ES6}{
  language=ES6,
  style=JSES6Base
}

\lstdefinelanguage{OManifest}{
  morekeywords=[1]{TITLE, DESCRIPTION, PIPELINE},
  morekeywords=[2]{false, null, true, boolean, number, undefined,
    Array, Boolean, Date, Math, Number, String, Object, ->},
  morekeywords=[3]{eval, parseInt, parseFloat, escape, unescape},
  sensitive,
  morecomment=[s]{/*}{*/},
  morecomment=[l]//,
  morecomment=[s]{/**}{*/}, 
  morestring=[b]',
  morestring=[b]"
}[keywords, comments, strings]

\lstdefinestyle{Manifest}{
  language=OManifest,
  style=JSES6Base
}

\definecolor{diffadd}{RGB}{0,128,0} 
\definecolor{diffdel}{RGB}{191,0,0} 
\definecolor{codebg}{RGB}{248,248,248} 
\definecolor{codeborder}{RGB}{220,220,220} 
\lstdefinelanguage{Diff}{ morecomment=[f][\color{diffdel}]-, morecomment=[f][\color{diffadd}]+, } 

\lstdefinestyle{diffstyle}{ 
    language=Diff, 
    basicstyle=\ttfamily\footnotesize, 
    backgroundcolor=\color{codebg}, 
    frame=single, 
    rulecolor=\color{codeborder}, 
    showstringspaces=false, 
    breaklines=true, 
    numbers=left, 
    numberstyle=\tiny\color{gray}, 
    xleftmargin=2em, framexleftmargin=1.5em,
    commentstyle=\color{gray},
    morecomment=[l]{//},
}
\setcopyright{popets}
\copyrightyear{YYYY}

\acmYear{YYYY}
\acmVolume{YYYY}
\acmNumber{X}
\acmDOI{XXXXXXX.XXXXXXX}
\acmISBN{}
\acmConference{Proceedings on Privacy Enhancing Technologies}
\begin{document}
%%
%% The "title" command has an optional parameter,
%% allowing the author to define a "short title" to be used in page headers.
% \title{OAuthWall: Mitigating OAuth Data Overaccess through a Local Firewall}
\title{OAuthHub: Mitigating OAuth Data Overaccess through a Local Data Hub}
% \title{OAuthHub: A Local OAuth Data Hub for Enforcing Principle of Least Privilege}

\newcommand{\sysname}{OAuthHub\xspace}

\newcommand{\qiyu}[1]{\textcolor{violet}{#1}}
\newcommand{\revise}[1]{\textcolor{blue}{#1}}
\newcommand{\haojian}[1]{\textcolor{red}{#1}}
\newcommand{\sssec}[1]{\vspace*{0.05in}\noindent\textbf{#1}}

\makeatletter
\def\ALG@special@indent{%
    \ifdim\ALG@thistlm=0pt\relax
        \hskip-\leftmargin
    \else
        \hskip\ALG@thistlm
    \fi
}
\newcommand{\Algorithm}{\item[]\noindent\ALG@special@indent \textbf{Algorithm:} }

\newcommand{\Initialization}{\item[]\noindent\ALG@special@indent \textbf{Initialize Parameters:} }
\makeatother

%%
%% The "author" command and its associated commands are used to define
%% the authors and their affiliations.
%% Of note is the shared affiliation of the first two authors, and the
%% "authornote" and "authornotemark" commands
%% used to denote shared contribution to the research.

\author{Qiyu Li}
\affiliation{%
  \institution{University of California San Diego}
  \city{La Jolla, California}
  \country{USA}}
\email{qiyuli@ucsd.edu}
\orcid{0009-0004-5174-9844}

\author{Yuhe Tian}
\affiliation{%
  \institution{University of California Riverside}
  \city{La Jolla, California}
  \country{USA}}
\email{yut009@ucsd.edu}
\orcid{0009-0008-5233-3647}

\author{Haojian Jin}
\affiliation{%
  \institution{University of California Riverside}
  \city{La Jolla, California}
  \country{USA}}
\email{haojian@ucsd.edu}
\orcid{0000-0001-5212-2235}

% \author{Ben Trovato}
% \authornote{Both authors contributed equally to this research.}
% \email{trovato@corporation.com}
% \orcid{1234-5678-9012}
% \author{G.K.M. Tobin}
% \authornotemark[1]
% \email{webmaster@marysville-ohio.com}
% \affiliation{%
%   \institution{Institute for Clarity in Documentation}
%   \streetaddress{P.O. Box 1212}
%   \city{Dublin}
%   \state{Ohio}
%   \country{USA}
%   \postcode{43017-6221}
% }

%%
%% By default, the full list of authors will be used in the page
%% headers. Often, this list is too long, and will overlap
%% other information printed in the page headers. This command allows
%% the author to define a more concise list
%% of authors' names for this purpose.
\renewcommand{\shortauthors}{}

%%
%% The abstract is a short summary of the work to be presented in the
%% article.
\begin{abstract}

Most OAuth service providers, such as Google and Microsoft, offer only a limited range of coarse-grained data access. 
As a result, third-party OAuth applications often end up accessing more user data than necessary, even if their developers want to minimize data access. 
We present \sysname, a development framework that leverages users' personal devices as the intermediary controller for OAuth-based data sharing between cloud services. 
The key innovations of \sysname are: (1) the insight that discretionary data access is largely unnecessary for most OAuth apps, which typically only require access at three well-defined moments—during installation, in response to user actions, and at scheduled intervals;
(2) a development framework that requires explicit declarations of intended data access and supports the three common access patterns through intermittently available personal devices; and (3) a centralized runtime permission model for managing OAuth access across providers.
We evaluated \sysname with three real-world apps on both PCs and mobile phones and found that \sysname requires moderate changes to the application code and imposes insignificant performance overheads. Our study with 18 developers showed that participants completed programming tasks significantly faster (9.1 vs. 18.0 minutes) with less code (4.7 vs. 15.8 lines) using \sysname than conventional OAuth APIs.

\end{abstract}
% local-hosted OAuth firewall that minimizes third-party access to users' data, which runs on users' devices locally, pre-processes users' data before relaying it to third-party applications, and denies malicious write requests. 

 % (Fig.~\ref{fig:oauth-exampls})



%
% Keywords. The author(s) should pick words that accurately describe
% the work being presented. Separate the keywords with commas.

%%
%% The code below is generated by the tool at http://dl.acm.org/ccs.cfm.
%% Please copy and paste the code instead of the example below.
%%

%% A "teaser" image appears between the author and affiliation
%% information and the body of the document, and typically spans the
%% page.
% \begin{teaserfigure}
%   \includegraphics[width=\textwidth]{sampleteaser}
%   \caption{Seattle Mariners at Spring Training, 2010.}
%   \Description{Enjoying the baseball game from the third-base
%   seats. Ichiro Suzuki preparing to bat.}
%   \label{fig:teaser}
% \end{teaserfigure}

% \received{20 February 2007}
% \received[revised]{12 March 2009}
% \received[accepted]{5 June 2009}

%%
%% This command processes the author and affiliation and title
%% information and builds the first part of the formatted document.
\maketitle

\section{Introduction}

% the details of data sharing between app developers (e.g., Uber) and service providers (e.g., Google), such as whether the app developers frequently pulling data and what data they get, are opaque to users~\cite{log_googleapi}. 

OAuth is an authorization protocol that enables delegated access to resources without requiring users to share their credentials~\cite{rfcoauth1.0, leiba2012oauth}.
For example, users can log in to applications (apps) like Uber and Zoom using their Google accounts and share their data, managed by Google, with these apps (Appendix ~\ref{sec:screenshots}).

This prevailing protocol faces two key privacy challenges. First, most OAuth implementations offer limited control and transparency to users. 
In a typical OAuth 2.0 implementation, the third-party OAuth app redirects the user to a service provider's login page to authenticate and grant consent. After that, the app receives an access token to interact with the service provider's resource server on the user's behalf. Once access is granted, data sharing occurs between third-party apps (e.g., Zoom) and service providers (e.g., Google), leaving users with limited control and transparency over when app developers request their data. 

To mitigate this concern, OAuth service providers started to allow users to access their personal data sharing under data protection law~\cite{Howtoacc61:online}. However, these efforts also face a few challenges.
First, many OAuth providers do not offer this access management feature to users. 
Second, most OAuth providers do not expire an app's access unless the app has not accessed users' data for six months~\cite{UsingOAu5:online,balash2022security, google:online}, and a few platforms never expire the access token~\cite{community_hubspot_2022}. 
Third, even if each service provider implements their own end-user review features, most users do not have the time to manage their granted accesses distributed across service providers~\cite{shen2021can, balash2022security,zufferey2023revoked}.

Second, most OAuth service providers offer only a limited range of coarse-grained data access. As a result, third-party OAuth apps often end up accessing more user data than necessary, even if developers want to minimize data access. 
For example, Uber's AwardWallet feature needs to extract users' itineraries from flight confirmation emails, but it has to request access to users' all email messages since it is the only applicable option in Gmail API.

For the second concern, existing solutions often require service providers to modify their server-side implementations~\cite{kroschewski2023save, cao2024stateful, RFC9396O83:RichAuth}. 
For example, Google Calendar API added a "free or busy" scope in addition to the all-or-nothing raw event scope~\cite{calendar:online}.
Lodderstedt et al. advocate a new protocol called OAuth 2.0 Rich Authorization Request~\cite{RFC9396O83:RichAuth}, which asks OAuth service providers to offer more granular access control mechanisms. 
Yet, implementing these protocol changes in practice is challenging~\cite{jain2014should}. Service providers need to adjust their APIs while ensuring backward compatibility and give long deprecation periods when changes are necessary.

\begin{figure*}[t]
  \includegraphics[width=1\linewidth]{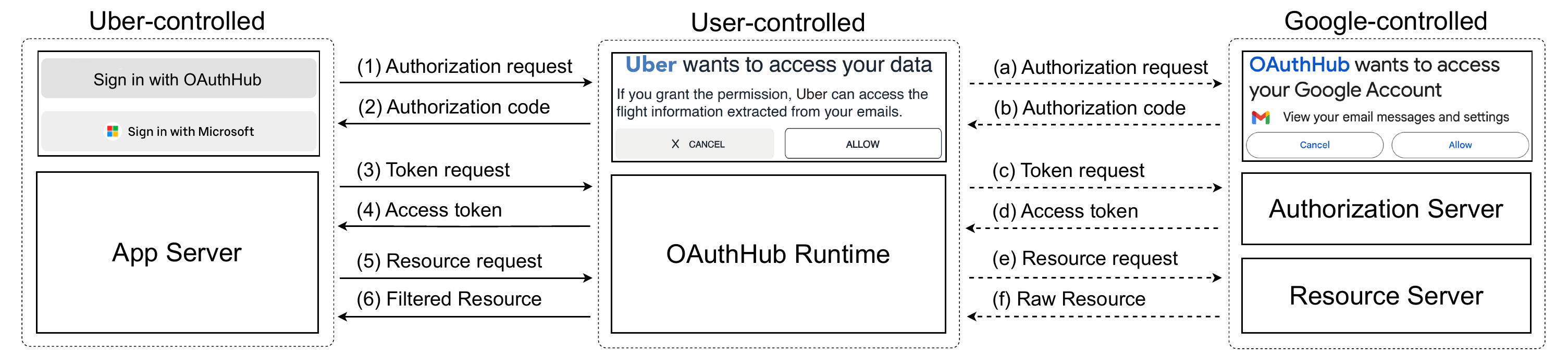}
  \vspace{-15pt}
  \caption{
  \sysname uses personal devices (e.g., smartphones, PCs) as an intermediary controller for data sharing between third-party apps (e.g., Zoom) and service providers (e.g., Google). \sysname retrieves data from service providers through OAuth (a-f), filters data locally, and allows third-party apps to access it through OAuth as well (1-6). 
  }
  \label{fig:oauth-wall-ui-flow}
  \vspace{-5pt}
\end{figure*}

This paper presents \sysname, a development framework that uses personal devices as the intermediary controller for OAuth-based data sharing between cloud services (Figure~\ref{fig:oauth-wall-ui-flow}).
\sysname uses personal devices (e.g., smartphones, PCs) as a data hub, which retrieves data from service providers (e.g., Google calendars) through OAuth (Figure~\ref{fig:oauth-wall-ui-flow} a--f), filters data locally, and allows third-party apps (e.g., Zoom, Uber) to access it also through OAuth (Figure~\ref{fig:oauth-wall-ui-flow} 1--6). 
In doing so, the local data hub allows app developers to reduce the data access without service providers' cooperation and offers users a centralized runtime permission model for managing OAuth access across providers~\cite{khandelwal2021prisec}.

While a few projects have explored data filtering to achieve data minimization~\cite{fernandes2018decentralized,jin2022peekaboo,cao2024stateful,chen2022practical}, the unique challenge for \sysname\ is leveraging users' intermittently available personal devices as the local data hub. \sysname introduces three key design ideas.

First, common OAuth implementations allow app developers to access users' data at their discretion, often without notifying users. One extreme example is that Uber Award Wallet may monitor users' emails continuously after the initial access request.
Our key insight is that \textbf{for most OAuth-based applications, granting developers discretionary access to user data is an unnecessary—and avoidable—privilege}. 
In analyzing 62 real-world OAuth apps, we found that most OAuth apps fall into three data access patterns: (1) {Install-time}, where the app only needs to pull data once to finish tasks like registering, linking, or setting up an account; (2) {User-driven}, where data access is initiated by the user's actions~\cite{roesner2012user}; and (3) {Scheduled time}, where the app needs to access data periodically even when the user is not actively using their service. 
We discuss the few apps that \sysname\ does not support in Section~\ref{sec:access_pattern}.

Second, common OAuth implementations require two servers: an authorization server, responsible for managing login requests and URL redirections, and a resource server, which handles API requests from third-party apps. 
Both servers must always remain on and have static IPs so that third-party apps can establish communication and redirect URLs. 
However, personal devices typically lack static IPs and are not consistently online due to power limitations and network conditions. 
Our insight into this challenge is that \textbf{servers running on intermittently available personal devices can support the three common access patterns if developers explicitly declare the desired access pattern.}
During both \textit{User-driven} and \textit{Install-time}, the user's device is already online, so the OAuth client can interact with the local device directly. In \textit{Scheduled time} scenarios, users have to power on their devices to access content updated when they are offline, so we can defer data access until the device is online again.

Third, the local data hub allows us to build a \textbf{centralized runtime permission model} for managing OAuth access across providers, similar to how Android manages App permissions today~\cite{micinski2017user}. Today, a user must grant Zoom full Google Calendar access (i.e., read and write) at installation time. In contrast, \sysname\ requires developers to explicitly declare the fine-grained access they need in a machine-readable manifest~\cite{androidManifestOverview,WhatisMU82:online,jin2022peekaboo} and allows users to selectively grant permissions to OAuth apps as well as impose constraints on the usage of resources, similar to Apex~\cite{nauman2010apex}. For example, users may only grant Zoom read access to their calendar at the installation time and then grant write access when they need to create a calendar event through Zoom later.

The design of \sysname\ also aims to minimize disruptions to the existing OAuth ecosystem. 
\sysname does not require service providers to change their server implementations. 
The only adjustments are that (1) third-party developers need to integrate a new login option, and (2) users need to install a mobile app on smartphones or a Chrome Browser Extension on PCs. 
We developed a lightweight library to ease the integration, allowing app developers to integrate \sysname\ with $\sim$50 lines of code. 
End users are motivated because it allows them to manage data access in a centralized manner and reduce unnecessary data disclosure~\cite{khandelwal2021prisec}. 
We implemented the prototype of \sysname\ on both PC and mobile phones and will open-source the framework.

We conducted detailed experiments to validate the design of \sysname. To understand the benefits and limitations of \sysname, we analyzed 218 OAuth apps collected from various marketplaces and examined how \sysname can mitigate data overaccess in these real-world use cases.
We evaluated \sysname with 18 students who play the role of developers and found that participants completed programming tasks more efficiently (9.1 v.s. 18.0 minutes) with less code (4.7 v.s. 15.8 lines) using \sysname than the conventional OAuth APIs.
We then recruited 100 participants on Cloud Research to test whether the fine-grained permissions enabled by \sysname can mitigate users' privacy concerns. Our results show that users were 56\% to 78\% less likely to deny the fine-grained permission requests through \sysname than the conventional OAuth permission requests. 
We developed three OAuth apps for PC and Android based on real-world scenarios by adapting sample source code provided by Google. Our experiments suggest that \sysname\ imposes modest performance overheads on users' local devices.

In this paper, we make the following contributions:
\begin{enumerate}[leftmargin=*,nosep]
    \item A development framework that uses individual personal devices as the intermediary controller for OAuth-based data sharing between cloud services.
    \item A characterization of when OAuth apps need data access. 
    \item A detailed prototype implementation and evaluation of \sysname\ on Android and PC. 
    % \item An open-source library for OAuth app developers to integrate \sysname.
\end{enumerate}

\section{Design Goal \& Threat model}
\label{sec:threat_model}
The key concept of \sysname\ is to use personal devices as the local data hubs to mediate the data sharing between OAuth service providers and third-party OAuth apps. 
Conventionally, these data-sharing activities are opaque to users and third-party auditors. 
By routing data through user-controlled devices and requiring developers to declare their data access in a machine-readable manifest, \sysname\ makes the process transparent, where app developers can assure users of their efforts in data minimization. It also enables a unified, centralized privacy management system (Section~\ref{sec:interface}).

There has been sustained interest in building hubs to enhance users’ privacy, leading to systems such as smart home hubs~\cite{jin2022peekaboo,tian2017smartauth}, network privacy hubs like Pi-hole~\cite{PiholeN58:online}, and privacy-setting hubs~\cite{khandelwal2021prisec}. \sysname extends this ecosystem by adding a new system primitive to hub developers’ toolboxes. \sysname shares many of the strengths and limitations of these hub-based approaches: it can improve users’ privacy by enabling data minimization and centralized data management, but third-party developers may be reluctant to adopt it because of incentives to monetize user data. 
We hope that \sysname\ can move the needle by offering two value propositions. First, it can increase the proportion of users who accept developers’ access requests; for example, Google’s Android team has found that exposing unnecessary information leads to unusually high denial rates~\cite{Permissi36:online,tahaei2023stuck}. Second, as hubs support a wider range of protocols, they become more useful to users, increasing the likelihood that users will adopt the hub model in practice.

\sssec{Threat model}. We envision that some well-intentioned third-party OAuth developers can incorporate a new login option powered by \sysname in their websites. \sysname's goal is to assure users that these developers only obtain data they declare and users agree to. 
We make the following assumptions:
(1) We assume that service providers (e.g., Google) will not disclose any personal data without appropriate credentials. Server-side data breach is an important problem, but orthogonal to the problem \sysname seeks to address. 
(2) We assume the user's local device is trusted. 
We conducted a detailed security analysis in Section~\ref{sec:security_analysis}.

% and showed that the risks of using \sysname\ are comparable to the risks inherent in conventional web-based OAuth implementations.

\sssec{System model}. Users need to run a local OAuth server on each of their devices where they want to use \sysname. Our current implementation supports both mobile and web-based apps. \sysname is implemented as a Chrome extension for web apps or as a mobile app for mobile devices. For web apps, the Chrome extension uses background scripts to listen for requests from the webpage, then filters data locally and forwards the requested data to the application.  \sysname requires network access to reach OAuth providers’ endpoints and communicates with client apps through the local environment, such as via a browser extension’s messaging APIs or a local service running on the user’s device.

\section{Understanding Data Access in OAuth Apps}
\label{sec:understand}

To gain insights into the OAuth data access problem, we conducted an analysis of 62 real-world apps that utilize OAuth APIs.
% from Google marketplaces.
% We also discuss how these analysis results inform the design of \sysname.

\sssec{Method}. 
We investigated apps on Google Workplace Marketplace (GWM), which commonly utilize Google's OAuth APIs. 
We first selected a diverse range of apps across all categories (e.g., creative tools, task management) in GWM, including both highly popular apps and niche ones.
We then manually installed the apps to identify the APIs and data types each app requested. 
We continued this exploration until we found no additional APIs or data types in the last ten apps.
Ultimately, we collected 62 OAuth apps that utilize Google's OAuth APIs (see Appendix~\ref{sec:applist}). 
We then analyzed what data these apps request, and what they actually need, assuming each app’s functionality based on its description. To complement this data analysis with actual data, we also studied a subset of 11 randomly selected apps using browser inspection and dynamic instrumentation tools (see Appendix~\ref{sec:oauthapps}).

 % to supplement our assumptions

Since the source code of the apps is not available, and since our analysis focuses on the use cases of OAuth apps rather than their actual behavior, we inferred the necessary data access scopes from the descriptions of their functionalities.
To mitigate potential bias, two authors independently labeled the results and cross-validated the labels, resolving any conflicts through discussion.

% While this inference may introduce bias, 

% 
\sssec{Results}. We make the following key observations. First, \textbf{most apps request more data access than they need to perform their tasks} (50 out of 62, 81\%). 
Most applications in our analysis require arbitrary read/write permission on the resource in their OAuth scopes, while they only need a small subset of the requested data (Appendix~\ref{sec:oauthapps}).
For example, 
% while Zoom requires events from the Google Calendar to assist users in scheduling, it only needs information about available time slots rather than detailed event data like location or name. 
% Further, several 
apps that interact with Google Drive often request extensive write permissions, while they only need to modify specific folders or file types (e.g., Notability). 
This finding suggests that the granularity of data access in existing OAuth APIs could be improved.

Second, \textbf{data access needs are diverse and app-dependent}. For example, Wellybox requires receipt attachments for reimbursement purposes, and TripIt needs transportation and lodging information in emails to organize travel itineraries (Table ~\ref{tab:data-overaccess-cases}). However, it would be challenging for service providers to implement and for developers to learn such a wide range of data granularities. Additionally, developers can minimize data access in many different ways, such as filtering only emails containing related information and removing unnecessary information in emails, while retaining the app's functionality. 
This finding motivates us to adopt a plugin architecture where developers can customize data access by chaining a set of operators.

\section{Personal Devices as the Data Hub}\label{sec:runtime}

% A key design of \sysname\ is to use a personal device as the intermediary controller for OAuth-based data sharing between cloud services. 
\sysname\ uses a personal device as the intermediary controller by establishing two OAuth-based data-sharing sessions: one between the developer and the local hub (Figure~\ref{fig:oauth-wall-ui-flow} 1-6), and another between the local hub and the service provider (Figure~\ref{fig:oauth-wall-ui-flow} a-f).

% By doing so, \sysname\ only requires third-party developers to modify their client code to integrate a new OAuth login option, without needing any cooperation from service providers.

Enabling the OAuth data-sharing session between the developer and the local hub (Figure~\ref{fig:oauth-wall-ui-flow} 1-6) is non-trivial. Conventionally, the OAuth protocol requires two key servers:
an authorization server, responsible for managing login requests and URL redirections, and a resource server, which
handles API requests from third-party apps. Both servers must
always remain on and have static IPs so that third-party apps
can establish communication and redirect URLs. However,
personal devices typically lack static IPs and are not consistently online due to power limitations and network conditions. 
This section describes how we leverage a new understanding of when OAuth apps need to request data to tackle this challenge. 

\subsection{When do OAuth Apps need Data?}
\label{sec:access_pattern}

Our key insight is that \textbf{allowing developers to request data at their discretion is an unnecessary privilege for most OAuth apps}. 
Conventionally, once a user grants access to third-party apps, data sharing occurs between the apps and the service providers, leaving users with limited control and minimal transparency. 
This design assumes that OAuth apps need to access user data at unpredictable times.

In analyzing 62 OAuth apps described in Section~\ref{sec:understand}, we found that most OAuth apps only need to request data at three types of times: 
(1) \textbf{Install-time} (7/62), where the app needs to pull data once to finish tasks like registering, linking, or setting up an account. For example, when using Quora, users can use OAuth to log in. 
(2) \textbf{User-driven} (36/62), where data access is initiated by users' actions.
For example, if a user edits a diagram in Draw.io, it would trigger Draw.io to save files to Google Drive. 
% Note that users' actions can be outside the OAuth app, such as booting a device.
(3) \textbf{Scheduled time} (19/62), where the app needs to access data periodically. 
For example, Lenovo Smart Frame needs to pull new photos daily to display them.

Fortunately, all these common OAuth data access can be supported by a personal device, even if the device is not designed to be always online. 
For both install-time and user-driven access, the access is triggered when users are actively using the devices, so the device is online.
While scheduled access may occur even when the personal device is not online, most OAuth apps can wait to fetch information until users actually need to use the service, so the device becomes online then. For example, a smart frame may remain off for several days and only needs to retrieve images when it is powered back on.

\sssec{What cannot be supported?}
We later expanded our search to more marketplaces (see Section~\ref{sec:eval_privacy}), actively seeking OAuth apps that cannot be supported by a personal device. We observed that an OAuth app requires discretionary access only when other users initiate actions that require access to a user’s OAuth-protected resources. 
For instance, the Jenkins-GitHub Plugin~\cite{Jenkins:online} triggers builds based on GitHub repository events, such as code commits, pull requests, and merges, which are retrieved through the OAuth protocol. 
Imagine that a collaborator might open a pull request while the original user is offline. Delaying the build process until the owner returns online is undesirable. \sysname does not support these OAuth apps; fortunately, cases that genuinely require discretionary access are rare (<3\%) among the apps we investigated.

 % that require discretionary access

% Since these ``social'' applications often do not involve sensitive personal data, they thus gain marginal privacy benefits from \sysname. Thus, we chose not to support these use cases.

% if its interactions are social and time-sensitive. 

% The Jenkins-GitHub Plugin may be triggered by a code commit from another user, so the owner's personal device may not be online. Further, the plugin needs OAuth data in a time-sensitive manner to prevent blocking other users. 
% So, these OAuth apps may access user data at unpredictable times. 
% \revise{\sysname does not support ``social'' apps where other users initiate actions that require access to a user’s OAuth-protected resources. For instance, in the GitHub CI scenario, a collaborator might open a pull request while the original user is offline. In such cases, delaying the build process until the user returns online is undesirable. Since these ``social'' applications often do not involve sensitive personal data, they thus gain marginal benefits from \sysname. So we chose not to support these use cases.}
% \sysname\ does not support these use cases that are both social and time-sensitive. 
% We found a few rare examples. 

% \vspace{-10pt}

\subsection{Declarative Access \& Declared Connections} 
\label{sec:communication_protocol}

We then made two key changes to the conventional OAuth implementations to support common OAuth access patterns using not-always-on users' devices and to enhance the control and transparency of OAuth-based data sharing. 
First, \sysname\ requires developers to explicitly declare when they will need to access user data through OAuth.
Second, instead of letting app servers initiate connections to the authorization/resource servers, \sysname\ lets the authorization/resource servers connect to the app server at times declared by developers. 
Note that the IETF OAuth framework~\cite{rfcoauth2.0} only specifies the data flow (e.g., access tokens, data access), but not the network flows. 
% So, we consider the modified implementation to be OAuth compliant. 

\sssec{Declared connections}. 
% Unlike servers that are publicly accessible, user devices predominantly operate behind Network Address Translation (NAT) barriers (e.g., wireless access points~\cite{grover2013peeking}), which block incoming connections. Because application servers cannot contact user devices behind firewalls or NATs, we let the local hub initiate the connection for data transfer. Specifically, for one-time and scheduled access, there is no server-initiated data request (Figure~\ref{fig:oauth-wall-ui-flow} step 5); instead, the hub pulls data from the service provider and posts the results to the developer-specified endpoint. 
By having third-party apps declare when they will need access, \sysname can establish connections on demand at the appropriate times. Because the hub runs on the user’s local device (often behind Network Address Translation barriers), external servers generally cannot initiate inbound connections to it. However, once an app declares its access timing, the hub can proactively initiate outbound connections to the resource server and the third-party app server as needed. For install-time access, the hub retrieves the data immediately after authorization and sends it to the app server. For user-driven access, the app client on the user’s device (e.g., a website or mobile app) sends a request to the hub at the moment of access, and the hub responds by delivering the requested data to the app server. For scheduled-time access, developers need to submit the access schedule to the hub during the authorization process.
This shifts the architecture from the server making trusted outbound calls to maintaining a public inbound listener. Consequently, developers must set up a public-facing API endpoint and implement robust DDoS protection, rate limiting, and signature verification to validate payloads from random IP addresses. In practice, common cloud endpoints (e.g., AWS) provide built-in DDoS protection and rate limiting configurations, while \sysname offers libraries to simplify payload signature verification. 

% the hub can proactively initiate outbound connections to the resource server and the third-party app server when needed

% the local hub initiates outbound HTTPS requests to deliver data to the application server.

\sssec{How to implement an \sysname\ client?}
Third-party apps can integrate \sysname\ by adding a new option, ``Sign in with \sysname'' (Figure~\ref{fig:oauth-wall-ui-flow}). 
We developed a lightweight TypeScript library to help third-party developers adopt \sysname (Figure~\ref{fig:lib}).
The library hides the complexity behind integrating \sysname, such as managing URL redirection and implementing API endpoints. 
Note that it is a common practice to implement OAuth clients using libraries due to the intrinsic complexity of handling access tokens.

\begin{figure}[htbp]
\vspace{-5pt}
\begin{lstlisting}[style=ES6, linewidth=0.45\linewidth]
//(1) generating authorization URL
const authorizationUrl = OAuthHubClient.generateAuthUrl({
    provider: service_provider, // e.g., google_calendar
    manifest: manifest,         // fine-grained data access
    redirect: redirect_url,     // callback URL after the OAuth
    accessType: access_type     // e.g., user_driven
});
//(2) exchanging authorization code for access token
OAuthHubClient.exchangeToken({
    provider: service_provider, // e.g., google_calendar
    manifest: manifest,         // fine-grained data access
    endpoint: endpoint_url,     // app server endpoint
    authCode: auth_code,        // auth code from callback URL
});
//(3) accessing or writing data through OAuthHub (user_driven access)
OAuthHubClient.query({
    provider: service_provider, // e.g., google_calendar
    manifest: manifest,         // fine-grained data access
    token: access_token,        // access token from step 2
    ...                         // other parameters
});
\end{lstlisting}
\vspace{-25pt}
\caption{The \sysname library provides APIs for: (1) generating authorization URLs, (2) exchanging access tokens, and (3) accessing data through \sysname. Notably, developers declare the desired access type in the initial API call (1).
}
\label{fig:lib}
\vspace{-10pt}
\end{figure}

\begin{figure*}[t]
  \includegraphics[width=1.0\textwidth]{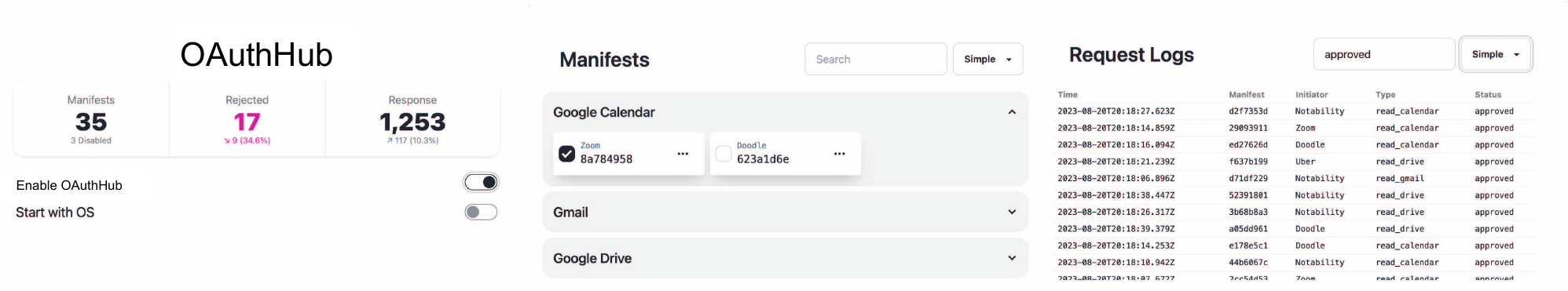}
  \vspace{-15pt}
  \caption{\sysname manifest management interface on the web. The global control panel (left) with the usage statistics of the runtime and toggles allows users to enable or disable \sysname. The manifests control panel (middle) can manage manifests across various services. The request logs interface (right) shows the access logs for user data via manifests for auditing.}
  \label{fig:oauthwall-ui2}
\end{figure*}

\noindent\textbf{How \sysname\ works?}
If a user taps ``Sign in with \sysname'', the login web page will redirect the user to a locally hosted authorization interface. 
After the user grants permission, \sysname\ will behave differently based on the developers' declaration of when they will need to access user data.

For install-time access, the local hub immediately sends the requested data to the API endpoint. For scheduled time access, the local hub sets up a background process to periodically send data as scheduled. If a device goes offline during a scheduled period, it will automatically reconnect to registered services once it is back online.
User-driven access is the only scenarios that require the API (3) in Figure~\ref{fig:lib}, where developers need to actively send requests to the local hub. For example, Draw.io may want to save a file through OAuth whenever a user edits a diagram.

The local hub will directly respond to a request if requested resources are readily available locally. If the authorization for \sysname is never granted or has expired, the user needs to go through a conventional OAuth authorization process with the resource provider (Figure~\ref{fig:oauth-wall-ui-flow} a-f). 
The local hub later stores all access tokens obtained from OAuth service providers, preventing the need for repeated user logins.

\section{Runtime Permission Model for OAuth}\label{sec:interface}

The local data hub also allows us to build a centralized permission model for managing OAuth access across providers.

% We will clarify the scope of our contribution and discuss what interaction features \sysname can enable to reduce the user burden.
% , similar to how Android manages App permissions today.

\subsection{On-demand Permission Granting}
\label{sec:user_control}

Current OAuth implementations often require users to grant all involved permissions upfront, similar to Android install-time permissions in the early days~\cite{wei2012permission}. 
For example, a user today must grant Zoom full Google Calendar access (i.e., read and write) at the installation time.

Inspired by the evolution of Android Permissions~\cite{felt2011android, felt2011effectiveness}, \sysname\ enables selective permission granting, similar to the modern Android permission model~\cite{peruma2018investigating,andriotis2016permissions,shen2021can}. With \sysname, Zoom needs to create separate manifests for read and write operations. Users can only grant reading access at the installation time and then grant writing access when they need to create a calendar event through Zoom for the first time.

This permission model is feasible because personal devices serve as intermediary controllers in OAuth-based data sharing. It would be challenging to enable this model in the conventional OAuth, because developers may request access while the user is offline.

 % each manifest to help users understand how their data will be processed

\begin{figure}[htbp]
  \includegraphics[width=0.48\textwidth]{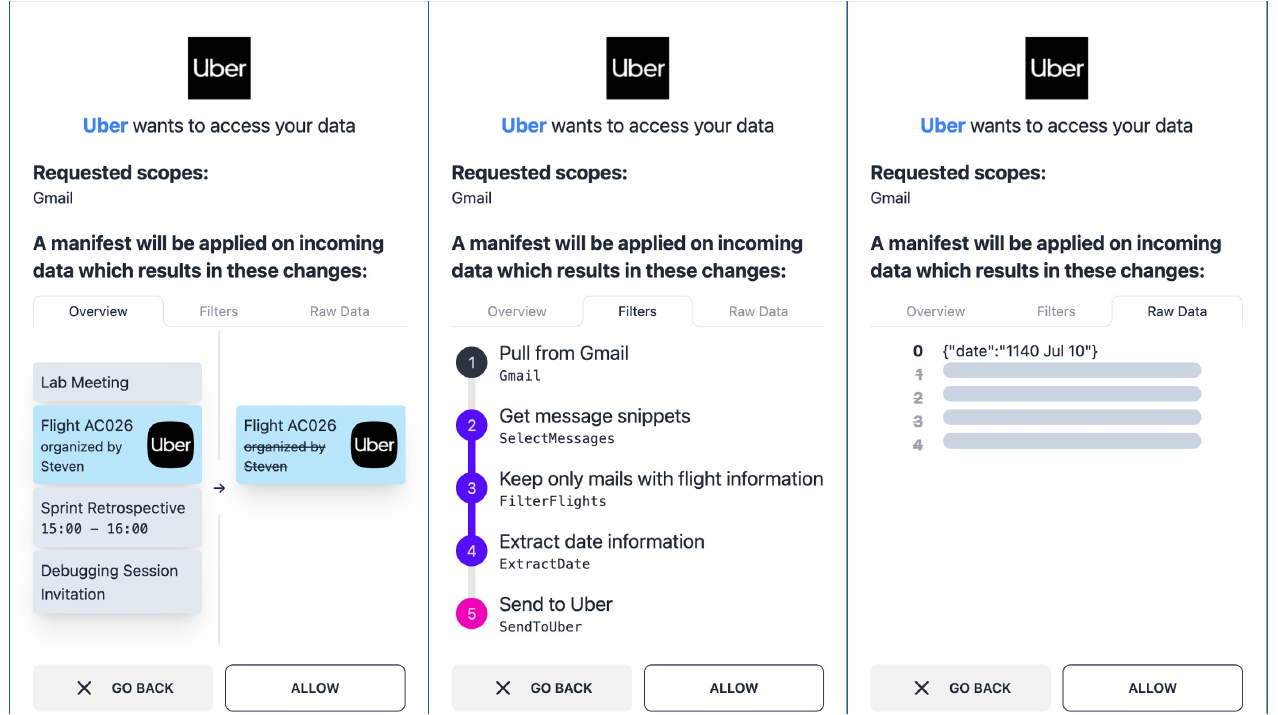}
  \caption{Example of an \sysname authorization interface on the Android device. \sysname runtime can generate centralized notice/control interfaces by parsing the machine-readable manifests: an overview (Left), the process steps (Middle), and a detailed data view (Right). 
  % renders the authorization interface based on the manifest, showing 
  }
  \label{fig:oauthwall-ui1-android}
  \vspace{-10pt}
\end{figure}

\sssec{Authorization interface}. \sysname\ generates an authorization interface by analyzing the manifest associated with an OAuth app (Figure \ref{fig:oauthwall-ui1-android}). The interface uses a multi-level design to balance detail and user comprehension.  It provides high-level summaries for average users, while technical users can inspect processing steps and preview the actual data transmitted. The interface comprises three views: an overview, processing steps, and a live preview. The overview displays both original and processed data. We use a static analyzer to generate step-by-step descriptions of the manifest pipeline, translating operator names and parameters into natural language. The live preview shows the actual data to be transmitted to third-party apps, based on the execution results of the manifest.

\begin{figure}[htbp]
    \hspace{-15pt}
    \includegraphics[width=\linewidth]{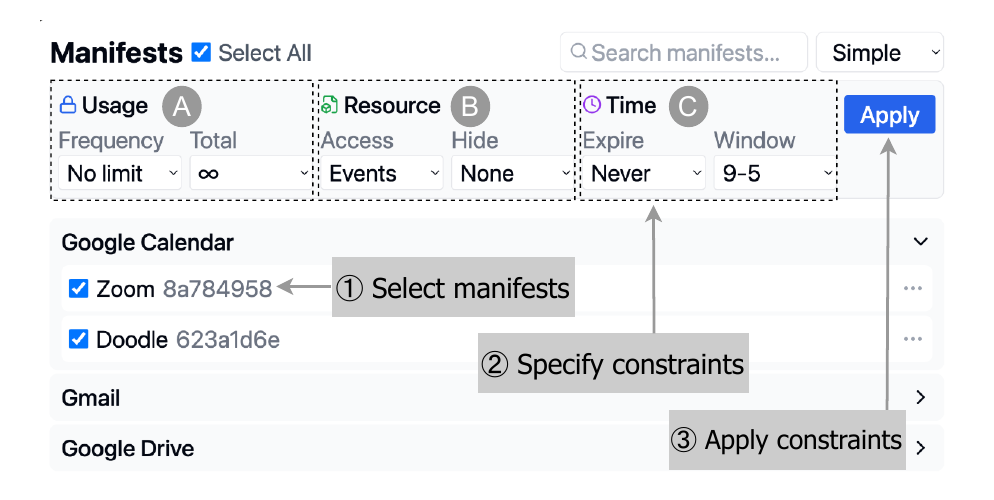}
    \vspace{-10pt}
    \caption{\sysname permission management interface on the web. \sysname allows users to specify three types of permission constraints through a centralized management interface: (A) \textbf{Usage constraints}: limit access frequency (e.g., twice per week) or total uses (e.g., one-time); (B) \textbf{Resource constraints}: restrict to specific resources (e.g., folders, file types) or obfuscate sensitive fields (e.g., names, emails); (C) \textbf{Time constraints}: set expiration duration (e.g., 24 hours) or restrict to specific time windows (e.g., business hours).}
    \label{fig:central_ui}
    \vspace{-10pt}
\end{figure}

\begin{figure*}[t]
  \includegraphics[width=1.0\textwidth]{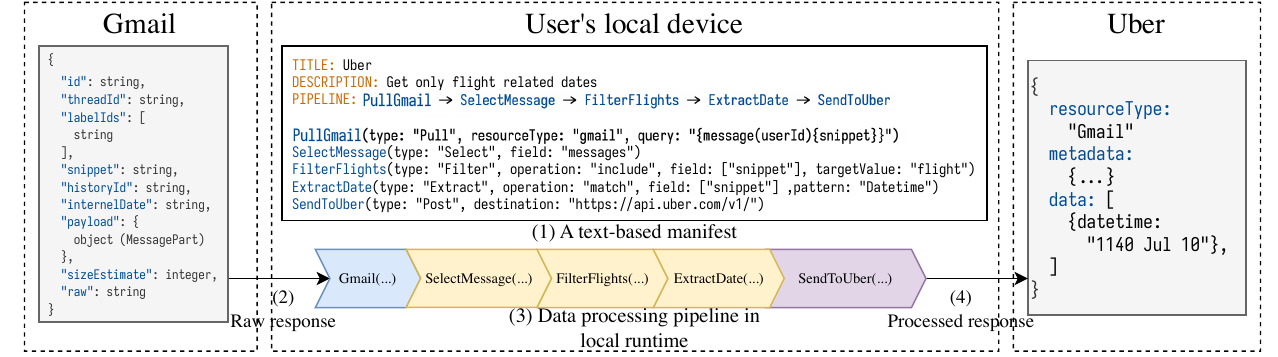}
\vspace{-15pt}
  \caption{
  Uber wants to scan users' emails for flight information to book rideshares upon arrival. \sysname can help Uber developers access Gmail data via OAuth while minimizing data access. Uber can create a manifest by chaining five operators (i.e., Pull, Select, Filter, Extract, and Post). When responding to Uber's data requests, \sysname (1) pulls raw data from Gmail, (2) filters for flight-related emails, (3) extracts the flight dates, and (4) sends only the necessary information to Uber.
}
\label{fig:oauthwall-arch-example}
\end{figure*}

\subsection{Centralized Privacy Management}
The local data hub can also enable an open platform around the OAuth protocol, where
well-intentioned third-party developers and privacy advocates can build centralized privacy management features. Because the manifests are machine-readable, \sysname can enable many privacy management features to enhance usability. We developed several features to demonstrate the feasibility.

% \qiyu{Interfaces shown are intended solely to demonstrate the feasibility of generating centralized notice/control interfaces by parsing manifests. Figures are illustrative prototypes and do not represent the final user experience.
% Because the manifests are machine-readable, \sysname can enable many privacy management features to enhance usability, such as privacy nutrition labels and auto-generated warnings. }

% We envision that third-party OAuth developers and privacy  adbocates will leverage the local data hub to build user-friendly interfaces to help users. This paper focuses on presenting the design of the open platform, while the exploration of the userinterface is deferred to future work.

Figure~\ref{fig:oauthwall-ui2} Right shows an interface where users can review the actual access logs of all OAuth apps. The logs contain the time, the initiator, the manifest, and the action type for each third-party request. These logs offer valuable insights into the data usage patterns of OAuth apps, benefiting both auditors and end-users. Figure \ref{fig:oauthwall-ui2} Middle shows another interface that allows users to revoke their permissions in a centralized manner, which displays a list of manifests they've authorized. Once revoked, \sysname expires the corresponding access tokens app developers obtained previously.
\sysname\ also allows users to create management policies across OAuth apps, including the number of access, the types of resources, and the time of access (Figure~\ref{fig:central_ui}).
Note that interfaces (Figure~\ref{fig:oauthwall-ui1-android},~\ref{fig:oauthwall-ui2},~\ref{fig:central_ui}) are illustrative prototypes to demonstrate the feasibility of automatically generating centralized notice/control interfaces by parsing manifests. They do not represent the final user experience.

\section{Fine-grained OAuth Data Access}
\label{sec:policy}

% Inspired by Peekaboo~\cite{jin2022peekaboo}, 
\sysname\ adopts a Pipe-and-Filter architecture~\cite{philipps99,jin2022peekaboo,privacystreams} to enable fine-grained OAuth data access.

\subsection{Operator-based Manifest}
\label{sec:manifest}

\sysname specifies the OAuth access as a data pre-processing pipeline connected with a set of stateless operators with known semantics, each performing a specific type of transformation. We design these operators to reduce OAuth data access and implement them as an open-source library. Third-party app developers declare data access by composing these operators into a pipeline and specifying the pipeline in a text-based manifest, which the local runtime uses to generate permission prompts and execute the pipeline using pre-loaded operator implementations locally. So developers can only interact with OAuth resources in ways supported by the filters. The operators are designed to support common data minimization only, rather than replicate all processing logic.

% Developers can configure the parameters of an operator to specify its behavior and save the pipeline as a text-based manifest. Given a manifest, \sysname\ assembles and executes a pre-processing pipeline using operator implementations pre-loaded locally. 

% \revise{The authors implement the filters as an open-source library. Third-party app developers declare data access by composing these filters into a pipeline and specifying the pipeline in a text-based manifest, which the local runtime uses to generate permission prompts.  So developers can only interact with OAuth resources in ways supported by the filters.  We design the filters to support common data minimizations only, rather than replicating actual processing.}

Figure~\ref{fig:oauthwall-arch-example} presents a pre-processing pipeline for Uber that extracts flight information from Gmail to offer recommendations based on the user's itinerary. 
The process begins with the ``PullGmail'' operator, which retrieves the user's Gmail content, followed by ``SelectMessages'' to extract relevant email messages. Next, ``FilterFlights'' uses regex matching to identify emails containing flight information, and ``ExtractDate'' extracts the date and time details. Finally, ``SendToUber" sends the data to the Uber service.

% \revise{The operator design is inspired by Peekaboo~\cite{jin2022peekaboo}: chaining a small set of common data filtering operators can enable a variety of data pre-processing tasks, which can significantly improve privacy.

% Similar to ~\cite{jin2022peekaboo,privacystreams}, 
\sysname operators are application-agnostic. When the \sysname loads JSON data from the service provider, the hub runtime parses the data according to the schema defined by the service provider. All OAuth data is represented in JSON, enabling filtering via a small set of SQL-like operators. In doing so, the same operators can interact with data streams from different providers.
Developers only need to adjust the operator properties to reuse them across applications. For example, a filter operator in Figure~\ref{fig:oauthwall-arch-example} is specified as [type: ``Filter'', operation: ``>'', field: ``start.dateTime'', targetValue: NOW]. Compared to SQL operators, these operators have well-defined semantics related to privacy.
% }

% % As in SQL, individual filters are defined with respect to a specific schema, while the operator set itself is schema-agnostic.
% Most data in OAuth data flows is tabular or list-based, allowing the operators to resemble SQL operators (e.g., SELECT, WHERE). 
% \qiyu{The authors implement the filters as an open-source library. Third-party app developers declare data access by composing these filters into a pipeline and specifying the pipeline in a text-based manifest, which the local runtime uses to generate permission prompts. All OAuth data is represented in JSON with a well-defined schema, enabling filtering via a small set of SQL-like operators. As in SQL, individual filters are defined with respect to a specific schema, while the operator set itself is schema-agnostic. We design the filters to support common data minimization only, rather than replicating actual processing.}

% converts the data into a standard schema for processing

\noindent\textbf{Operators v.s. Attributes}. One alternative design to implement fine-grained data access is through attribute-based data access~\cite{hu2015attribute}. For example, \sysname\ may introduce a declarative attribute-based data access, such as flight confirmation email access, to handle the Uber scenario. We decided to use the operator-based approach because it can enable more flexible data access management. 
% is more effective in achieving diverse access reduction. 
For example, \sysname may further reduce the access in Figure~\ref{fig:oauthwall-arch-example} by inserting operators that remove attachments from the email and obfuscate names in the email text. In contrast, it would be onerous for service providers to implement numerous potential attribute-based data accesses and for developers to learn these APIs.

\subsection{Operator Design}
\label{sec:operator}
% While the exact desired data transformations vary across use cases, the data pre-processing actions have similar high-level semantics. For example, a \textit{Filter} transformation chooses a smaller portion of the original data by imposing constraints based on keywords or pattern matching, such as filtering email with upcoming flight information, screening Zoom events in Google Calendar, or selecting specific files in Google Drive.

% \sysname\ uses a predefined set of operators that can be chained to achieve the desired data actions. 
We used best practices in API design~\cite{bloch2006design} to guide the design of these operators.
% We used an approach similar to ~\cite{jin2022peekaboo} and  to design the operators. 
We started with operators in ~\cite{jin2022peekaboo,privacystreams} and used them to implement the use cases of OAuth apps collected in Section \ref{sec:understand}. We then iteratively expanded the operator design to accommodate additional use cases. Finally, we removed the unused operators. This process resulted in thirteen operators grouped into five categories: \texttt{provider}, \texttt{reduction}, \texttt{transform}, \texttt{network}, and \texttt{utility} (Appendix ~\ref{sec:operators}). 

\sssec{Data exchange}. Since \sysname serves as a proxy for processing raw data requests, it needs to mediate data between OAuth resource servers and third-party applications. For OAuth apps retrieving data from OAuth resource servers, we provide the ``Pull" operator to fetch only the necessary data. We allow developers to specify exactly what data they need from an OAuth resource using a GraphQL query, a widely-used data query language for declarative API fetching~\cite{hartig2018semantics}. For example, in Figure \ref{fig:oauthwall-arch-example}, the ``Gmail" operator specifies only the snippets of email messages are needed. This approach offers three benefits. First, it reduces the risks of data overaccess and improves network efficiency by transmitting only the necessary data. Second, it enhances transparency by exposing what specific data is being used. Third, it reduces the burden on developers to adapt to new API versions, as the data representation tends to remain compatible across API evolution~\cite{fokaefs2011empirical}. Developers can use the \textit{Post} operator to send the processed data to their application services.

To allow apps to perform actions on the OAuth resource server, we provide \textit{Receive} operator to handle requests from third-party applications. We require developers to encode their requests in JSON to maintain uniform data representation with OAuth data. This includes an \texttt{action} field representing the API name of the OAuth resource (e.g., create, update, delete), a \texttt{body} field for the HTTP request body (e.g., file content), and a \texttt{parameters} field listing all parameter names and values (e.g., metadata). The \textit{Write} operator acts as a wrapper, invoking the API of the OAuth resource with the specified parameters to perform the actions.

% For OAuth apps requiring data retrieval from OAuth resource servers, we offer the \textit{Pull} operator to fetch only the necessary data. It requires developers to specify the resource type and exactly what data they need using GraphQL queries, a widely used data query language that enables declarative fetching from APIs. For example, in Figure \ref{fig:oauthwall-manifest-execution}, the "Gmail" operator specifies only the snippet field in email messages is needed. This approach offers two benefits. Firstly, it helps developers to offer transparency by exposing what data they need to use. Secondly, this data-centric approach aids in maintaining compatibility across API evolutions. While OAuth APIs may evolve, the data representation tends to remain relatively stable, reducing the burden on developers to adapt to new API versions for data retrieval methods. After the data processing, we provide \textit{Post} operator to send the processed data to the OAuth app server.
% Each data source corresponds to a specific OAuth scope ands provide raw data encoded in JavaScript Object Notation (JSON) \cite{JSON:online}.  

\sssec{Data filtering and transformation}. 
Due to the coarse nature of OAuth scopes, OAuth data often includes irrelevant information for the third-party application's needs. For example, Uber only needs emails with flight information rather than all emails. While developers can minimize data by specifying necessary fields using GraphQL, we also provide \texttt{Reduction} and \texttt{Transform} operators to support more complex logic. The \texttt{Reduction} operators trim data to only preserve the necessary information. For example, \textit{Filter} operator can be used to exclude irrelevant data based on specific conditions, such as removing emails unrelated to flights. \textit{Extract} operator can retrieve information based on particular data patterns, like extracting flight dates from related emails. The \texttt{Transform} operators process the data to produce derived or aggregated results. \textit{Aggregate} operator combines multiple data items to generate statistical reports, such as counting the number of responses for a specific Google form. \textit{Map} operator transforms individual data items, such as converting birth dates into age categories for demographic analysis.
\textit{Anonymize} operator modifies data to enhance privacy, like adding noise to ages in user profiles.

% Therefore, the process of narrowing the requested data is one major method to enforce data minimization in the pipeline, which is done by operators like \textit{Filter} and \textit{Limit}. 
% These dataset narrowing operators receive a set of predefined conditions and a list of fields they need to narrow as their parameters.
% NOTES: Filter, Sort and Limit Operators

% Developers need to specify the OAuth scope requested by the application as a parameter to the corresponding operator. 

% An example of dataset narrowing may be used in Uber Travel. 
% In this case, the application needs to access the user's Gmail data to find emails related to flight times and send recommendations to users according to their itinerary. 
% Currently, this is done by using the access token granted by the user to pull all recent emails in the user's account and perform the filtering on Uber's servers. 
% Unfortunately, this violates the data minimization principle. 
% Moreover, end users and other experts cannot reliably verify that Uber only used the emails to retrieve flight itineraries.
% With the help of \sysname, Uber can author a manifest indicating that they only require emails with certain keywords to be included.

\sssec{Specifying action constraints}. Since action requests are also represented in JSON format, well-intended developers can use the \textit{Filter} operator to restrict their actions similarly. For example, they can limit file paths for write operations by filtering requests where the file's parent ID matches the destination folder's file ID. If there is no match, the \textit{Write} operator will not execute any action with null input. This additional sanity check helps mitigate the risks of data overaccess and maintain data integrity.
\section{Case Studies}
\label{sec:case_study}
We present three example \sysname\ use cases.

% \qiyu{We will add a figure to illustrate the changed lines and explain the integration workflow.}

% We implemented three end-to-end applications to assess the expressiveness of \sysname's manifest language and its fitness for real-world scenarios. 

\sssec{Zoom accesses Google Calendar}. Zoom's current integration with Google Calendar requires users to grant Zoom permission to view, edit, share, and permanently delete all the calendars they can access. 
\sysname\ can help Zoom mitigate this over-privilege access using a manifest that filters calendar events that do not contain Zoom meeting links. 
We implemented the client using the example code provided by the Google Calendar API. We then added 44 lines of code and removed 20 lines to support \sysname. 

% \begin{lstlisting}[style=Manifest,label=lst:zoom]
% TITLE: Zoom
% DESCRIPTION: Get all upcoming Zoom meetings
% PIPELINE: PullCalendarEvents -> SelectEvents -> FilterTime 
%                              -> FilterZoom -> PostToZoom

% PullCalendarEvents(type: "Pull", resourceType: "google_calendar", 
%                    query: "{ events(calendarId) {...} }")
% SelectEvents(type: "Select", field: "events")
% FilterTime(type: "Filter", operation: ">", 
%            field: "start.dateTime", targetValue: NOW)
% FilterZoom(type: "Filter", operation: "match", 
%            field: ["location", "description"], 
%            pattern: "Zoom URL", requirement: "any")
% PostToZoom(type: "Post", destination: "www.zoom.us")
% \end{lstlisting}
% \vspace{-20pt}

\sssec{Uber accesses Gmail content}. Uber wants to access users' flight confirmation emails to offer travel personalization. Its current integration with Gmail requires users to permit Uber to read all their emails and settings. 
\sysname\ can help Uber mitigate this overaccess by extracting only the relevant flight date information using the manifest in Figure~\ref{fig:oauthwall-arch-example}. 
% Note that while this manifest is designed for English users, it can be easily extended to other languages by adding relevant keywords for email filtering.
We implemented the client using the example code provided by Gmail API. We then added 45 lines of code and removed 28 lines to support \sysname.

% \begin{figure}[htbp]
% \begin{lstlisting}[style=Manifest, label=lst:notability]
% TITLE: Uber
% DESCRIPTION: Get only flight related dates
% PIPELINE: Gmail -> SelectMessage -> FilterFlights 
%           -> ExtractDate -> SendToUber
% Gmail(type: "Pull", resourceType: "gmail", 
%       query: "{message(userId){snippet}}")
% SelectMessage(type: "Select", field: "messages")
% FilterFlights(type: "Filter", operation: "include", 
%               field: ["snippet"], targetValue: "flight")
% ExtractDate(type: "Extract", operation: "match", 
%             field: ["snippet"] ,pattern: "Datetime")
% SendToUber(type: "Post", destination: "https:api.uber.com/v1/")
% \end{lstlisting}
% \vspace{-2em}
% \caption{Uber-Gmail integration}
% \label{fig:notability}
% \end{figure}

\sssec{Notability writes to Google Drive}. 
Notability is a note-taking application that combines handwriting, photos, audio and typing in a customizable interface. 
Notability's backup feature requests permissions to read, write and delete files in all the folders on Google Drive. 
\sysname\ can prevent Notability from writing to arbitrary paths in Google Drive by filtering the parameters of requests. 
We implemented the client using the example code provided by Google Drive API. We then added 51 lines of code and removed 36 lines.

% \begin{lstlisting}[style=Manifest, label=lst:notability]
% TITLE: Notability
% DESCRIPTION: Backup notes to Google Drive
% PIPELINE: ReceiveRequest -> FilterPath -> Upload

% ReceiveRequest(type: "Receive", source: "www.notability.com")
% FilterPath(type: "Filter", operation: "match", field: ["parents"], 
%            targetValue: "folderId")
% Upload(type: "Write", action: "create", resourceType: "google_drive")
% \end{lstlisting}
% \vspace{-2em}
% \label{fig:notability}

% \begin{figure}[htbp]
% \begin{lstlisting}[style=Manifest, label=lst:notability]
% TITLE: Notability
% DESCRIPTION: Backup notes to Google Drive
% PIPELINE: ReceiveRequest -> FilterPath -> Upload

% ReceiveRequest(type: "Receive", 
%                source: "www.notability.com")
% FilterPath(type: "Filter", operation: "match", 
%            field: ["parents"], targetValue: "folderId")
% Upload(type: "Write", action: "create",
%        resourceType: "google_drive")
% \end{lstlisting}
% \vspace{-2em}
% \caption{Notability-Google-Drive integration}
% \label{fig:notability}
% \end{figure}
\section{Implementation}\label{sec:implementation}

The development framework of \sysname comprises a developer-facing library and a local hub runtime.

\sssec{Developer-facing library}. We developed a lightweight TypeScript library ($\approx$ 300 lines of code) to help third-party OAuth app developers integrate \sysname. 
The library hides the complexity of implementing OAuth protocols, ensuring that developers only have to make minimal code modifications. To support \sysname, developers only need to generate redirect URLs and call APIs to retrieve and consume access tokens (Figure~\ref{fig:lib}).

We also developed an Integrated Development Environment (IDE) to assist developers in authoring and debugging the manifest. Our IDE features syntax highlighting, auto-completion, and built-in documentation. Developers can insert a \textit{Debug} operator in the manifest pipeline to output intermediary results for debugging.

\sssec{Local hub runtime}.  We developed a hub runtime to enable the permission model on users' personal devices. 
The hub runtime executes operator-based manifests, hosts local-hosted authorization web pages, manages access tokens from OAuth service providers, and offers data access management interfaces. 
We implemented the core runtime using Node.js, and then developed a Chrome extension for PCs and an Android app for mobile phones.
Since Node.js is cross-platform, we reused most of the code across PC and mobile.

\noindent \textbf{Security}. One potential concern is that local devices may become lucrative targets for attackers, as \sysname\ centralizes the management of OAuth tokens. To mitigate this risk, \sysname leverages platform-specific security mechanisms. For Chrome extensions, \sysname invokes the \texttt{chrome.identity.getAuthToken} API for authorization, which eliminates the need to store the client secret within the extension, maintains tokens in Chrome's internal secure storage, and automatically obtains and refreshes access tokens. For Android, \sysname uses the AppAuth SDK~\cite{appauth:online} to perform OAuth authorization with servers, which follows OAuth security best practices for native apps~\cite{rfcoauthnative}. To reduce the risk of a single point of compromise across multiple services, \sysname encrypts tokens using unique keys for each service and stores the keys in secure OS storage such as Android’s KeyStore~\cite{KeyStore:online}.

\section{Evaluation}\label{sec:eval}

This section presents detailed experimental evaluations of \sysname, including its security analysis (Section \ref{sec:security_analysis}), effectiveness in mitigating overaccess risks in OAuth apps (Section \ref{sec:eval_privacy}),
 usability for developers (Section \ref{sec:eval_developer}), how it addresses users’ privacy concerns (Section \ref{sec:eval_user}), and system performance (Section \ref{sec:eval_performance}). 

\subsection{Security Analysis}\label{sec:security_analysis}
Using personal devices as intermediaries introduces additional attack surfaces. 

\sssec{User identity impersonation}. Since \sysname lets third-party apps communicate with the user-controlled device rather than directly with the authoritative OAuth provider, third-party apps need to verify the user’s identity during authorization and in subsequent communications to prevent false claims of another user’s account. \sysname obtains identity tokens (e.g., JWTs) from the service provider, and applications verify these tokens with the provider as usual to confirm user identity. As a result, \sysname cannot falsely claim access to an unrelated account, because the tokens cryptographically bind the user to the account. During authorization, \sysname generates a unique key pair and shares the public key with the app; it then signs all subsequent payloads with the private key, enabling the app to verify authenticity from random IP addresses.

\sssec{Physical device loss/theft}. A concern is that users could lose their personal device, allowing an unauthorized person with physical access to access all users' data across all OAuth service providers. \sysname\ employs an on-demand loading strategy: tokens and data are stored only when explicitly requested and automatically expire according to the session policies of service providers. 
% As a result, the risks of \sysname\ are comparable to the risks inherent in conventional web-based OAuth implementations.

\sssec{Local device compromise}. 
Attackers may steal \sysname tokens by compromising the user devices (e.g., rooting the OS). 
\sysname mitigates this by only storing encryption keys and encrypted tokens in platforms' secure storage.

\sssec{Man-in-the-middle attacks}. Malicious third-party libraries or apps may intercept tokens or authorization data in transit. \sysname mitigates this by implementing PKCE~\cite{rfcpkce}, which requires a matching code verifier for token exchange to prevent authorization code interception attacks. Additionally, \sysname verifies web page origins and app signatures to ensure request integrity, and rotates access tokens (i.e., issuing a new token after each access and invalidating the prior one) to prevent token replay attacks.

\sssec{UI Spoofing}: Malicious apps may impersonate \sysname to trick users into granting access. \sysname mitigates this by performing authorization through platform’s secure UI framework (e.g., \texttt{chrome.identity.launchWebAuthFlow}), ensuring users interact only with system-rendered consent screens.

\sssec{Users do not read the interface}. Abusive developers may request OAuth data excessively without filtering (e.g., fetching raw data every second), and users may approve without reading the interface. Because manifests are machine-readable, future work can develop automated algorithms that analyze access patterns, detect malicious or overly aggressive behavior, and potentially auto-deny such requests.

\sssec{Security benefit}. One benefit is that \sysname manages refresh tokens on behalf of developers, which may prevent developers from mistakenly exposing refresh tokens in the front-end code, as found in prior studies~\cite{zhang2023don, shi2025skeleton}.

\subsection{Mitigating Overaccess in OAuth Apps}
\label{sec:eval_privacy}
We examined how \sysname mitigates the overaccess risks in 218 OAuth apps qualitatively.

\sssec{Method}. 
We investigated a broader set of  OAuth apps to assess how well \sysname generalizes to different use cases.
We first collected 21 OAuth app marketplaces and grouped them into four categories: platform, productivity, communication, and programming. We then selected the most popular marketplaces for each category: Google, Trello, Slack, and Github. For each marketplace, we selected the apps similarly to the study in Section~\ref{sec:understand}. This search concluded with 218 apps in total. Two authors then collaboratively implemented a manifest for each app function to explore ways to mitigate data overaccess. We then independently and manually coded the manifest and discussed high-level categories for the resulting codes to agree on a selected scheme.

% To validate the feasibility of \sysname, we created manifests to cover the use cases outlined in \S \ref{sec:understand}.

% Each OAuth app can serve multiple functions. For example, Zoom needs to view and schedule meetings on Google Calendar. Two authors collaboratively implemented manifests for each function, exploring methods to restrict OAuth access. This process resulted in XX manifests.

\sssec{Results}. In our analysis of 218 apps, only six apps require discretionary access, which our system cannot support. These apps fall into three categories: (1) CI pipeline triggers (e.g., Jenkins, Google Cloud Build) that trigger builds when code changes occur, (2) Security scanning (e.g., GitGuardian) that detects exposed secrets for each pull request, and (3) Task sync tools (e.g., Gitlab) that maintain consistency across project management platforms across team members. For the remaining apps, we examined how \sysname minimizes their data access.
 We created a codebook with 9 common methods for mitigating data overaccess, grouped into three categories: data minimization, action restriction and access timing restriction (Table \ref{tab:eval_coverage}). 
 % Our investigation revealed that 117  apps ask for unnecessary permissions.
 % often due to coarse OAuth scopes
 % This can be addressed by \sysname through fine-grained permissions. 

\sysname provides four data minimization methods when retrieving data from OAuth resource servers: data filtering, content selection, data anonymization, and data aggregation. Data filtering is the most common method (161 out of 218 apps), which screens related data items based on specified criteria, such as file types or app-specific tags. Content selection is also very common (139/218). Since many apps only need specific types of information in the data, we can extract only necessary information from complex OAuth data representations, such as retaining only related information (i.e., localId, email and language) within user profiles for Google Colab. Data anonymization (51/218) enhances user privacy when the apps need to access Identifiable Information (PII), such as hashing user emails. 
Data aggregation is currently available for only 7 out of 218 apps. Data aggregation is applicable to only 9 out of 218 apps, as the remaining apps require raw or individual-level data. We currently support 7 of those 9 apps, as the remaining two depend on complex aggregation operations (e.g., conditional sums and nested aggregations) that our system does not yet support. We anticipate our supported operations will expand over time to accommodate more use cases.

% Data aggregation is currently available for only 7 out of 218 apps due to limited supported operations, but 

For apps interacting with OAuth resources, \sysname offers two common types of action restriction: limiting action types and constraining action conditions. The first (176/218) controls which API methods third-party apps can use, like allowing only updates to documents but not creation or deletion in Google Docs. The second (155/218) imposes specific constraints on actions, such as limiting file paths or accepted file types for upload.

\sysname also enforces access timing restrictions based on data access patterns. For install-time apps (9/218), it only allows one-time use after installation. For user-driven apps (135/218), background access is not feasible when the user's device is offline. For scheduled time apps (68/218), it limits access duration and frequency. 

% \revise{\sysname does not support time-sensitive scheduled tasks when devices are offline. Deferring tasks is not acceptable for two apps identified, such as social apps that require immediate action.}

\begin{table*}[htbp]
    \centering\small
    \begin{tabular}{|l|l|l|c|}
    \hline
    Category & Method & Usage Scenario & \# Support \\
    \hline
    \hline
    \multirow{4}{5em}{Data Minimization}  & Data Filtering & Apps require only a portion of data that is relevant & 161 / 218 \\\cline{2-4}
          & Content Selection & Apps only need partial data in specific fields & 139 / 218 \\\cline{2-4}
         & Data Anonymization & Apps require Personally Identifiable Information (PII)
 & 51 / 218 \\\cline{2-4}
         & Data Aggregation & Apps only need aggregated results & 7 / 218 \\
    \hline
    \multirow{2}{5em}{Action Restriction} & Restricting Action Type & Apps only need to perform specific types of operations & 176 / 218 \\\cline{2-4}
         & Restricting Action Conditions & Apps only need to perform actions satisfying specific conditions
& 155 / 218 \\
    \hline
    \multirow{3}{5em}{Timing restriction}
    % Downscoping Permissions & Apps only need to access resources used with the app & 136 / 207 
         & Allowing only once  & Apps are intended for one-time use during installation & 9 / 218 \\\cline{2-4}
         & Allowing only while using  & App only need access when user is actively using the app & 135 / 218 \\\cline{2-4}
         & Allowing only at scheduled times  & Apps need to access data periodically & 68 / 218 \\
         % & & & / \\
    \hline
    Not supported & \centering -- & Apps require discretionary access & 6 / 218 \\
    \hline
    \end{tabular}
    \caption{We implemented manifests for 218 OAuth apps, analyzed the types of methods for mitigating data overaccess in these manifests, and examined the applicable usage scenarios.}
    \label{tab:eval_coverage}
\end{table*}

\subsection{API Usability for Developers}
\label{sec:eval_developer}
We conducted an IRB-approved study with 18 developers to evaluate the usability of our API.

\sssec{Method}. We aimed to study how \sysname can support developers to access and process personal data. Specifically, we are interested in (1) whether \sysname is faster and easier to use than the conventional OAuth APIs for app developers and (2) whether \sysname helps developers effectively minimize data access.

We asked participants to complete programming tasks involving personal data. We derived four tasks from common OAuth app use cases in Section \ref{sec:understand} (see Table \ref{tab:developer_task}). The study was a within-subjects design, where participants used both conventional OAuth APIs and \sysname to complete two tasks each. We counterbalanced the presentation order of APIs and tasks across participants. We did not disclose that one of the APIs was developed by the authors to prevent introducing potential bias.

Since CS students are considered acceptable substitutes for developers~\cite{tahaei2022recruiting},
we recruited 18 CS students from multiple universities. Participants in the study (13 identified as male, 5 identified as female, aged 19-25) were undergraduate and graduate students with at least 3 years of programming experience. Each participant received a 20 USD gift card as compensation for their time. The study was performed in our lab or remotely over the Zoom meeting. We obtained participants’ consent electronically via Google Forms at the beginning of each study.

The study took about 90 minutes for each participant, including two 45-minute sessions. Each session included a 10-minute walk-through, a 30-minute programming period and a 5-minute debriefing. During the walk-through period, we guided participants through warm-up tasks to get them familiar with \sysname or conventional OAuth APIs. In the programming period, we instructed participants to complete the assigned tasks.  They were free to search the Internet for documentation and solutions. For a fairer comparison, we did not require the participants to declare the scopes using conventional
OAuth APIs, as \sysname automatically requests and manages OAuth scopes. During the debriefing period, participants completed the NASA Task Load Index (NASA-TLX) questionnaire~\cite{hart2006nasa} to rate six aspects of cognitive load on a 7-point Likert scale. After the study, participants participated in a semi-structured interview to discuss their experiences using two APIs.

\begin{table}[htbp]
    \centering\small
    \vspace{-5pt}
    \begin{tabular}{|p{3.5em} p{4.5em}  p{18em}|}
    \hline
    Task ID & App & Task description \\
    \hline 
    \hline
    Task 1 & Zoom & Get all upcoming Zoom meetings from Google Calendar\\
    \hline
    Task 2 & WellyBox & Extract financial receipt attachments from Gmail \\
    \hline
    % Task 3 & Zotero & Edit citations in Google Docs \\
    Task 3 & FormLimiter & Get the number of valid responses from Google Forms \\
    \hline
    Task 4 & Notability & Backup notes to Google Drive\\
    \hline
    \end{tabular}
    \caption{The programming tasks used in the developer study. 
    % These tasks were crafted based on common use cases of popular apps. In the study, we provided participants with a detailed task description and the expected output.
    }
    \label{tab:developer_task}
    \vspace{-23pt}
\end{table}

\sssec{Results}. Table \ref{tab:developer_study} shows the results of our developer study. All participants completed the tasks successfully for both APIs. Regarding the developers' efficiency, participants spent less time (average of 9.1 v.s. 18.0 minutes) to complete the tasks using \sysname than using the conventional OAuth APIs. The Holm-Bonferroni corrected Mann-Whitney U test showed significant differences in completion time between the two APIs (p $<$ 0.05 for all tasks). We also counted the lines of code (LOC)  in the final programs. On average, participants used less code (4.7 v.s. 15.8 lines) to complete the tasks with \sysname than the conventional OAuth APIs. Participants could finish tasks 1 to 3 using \sysname with only 5-7 lines of code, and task 4 in just 2 lines.

To quantify the privacy benefits of \sysname, we tested the manifests created by participants using a real-world dataset curated from the authors' personal data, to avoid collecting sensitive data from participants. The dataset consists of 500 emails, 97 calendar events, and 9 form responses. We measured the percentage of data egress reduction in JSON entries and bytes. On average, \sysname filtered out 96\% of emails, 97\% of calendar events, and 44\% of form responses, resulting in data reduction rates of 95\% for Zoom and over 99.9\% for both WellyBox and FormLimiter. The observed data reduction results from data pre-processing on the edge. For example, Zoom previously retrieved all raw Google Calendar events. In contrast, \sysname filters data locally on users’ devices and only relays the processed results.

\begin{figure}[htbp]
    % \centering
    % \hspace{-1.6em}
    \includegraphics[width=\linewidth]{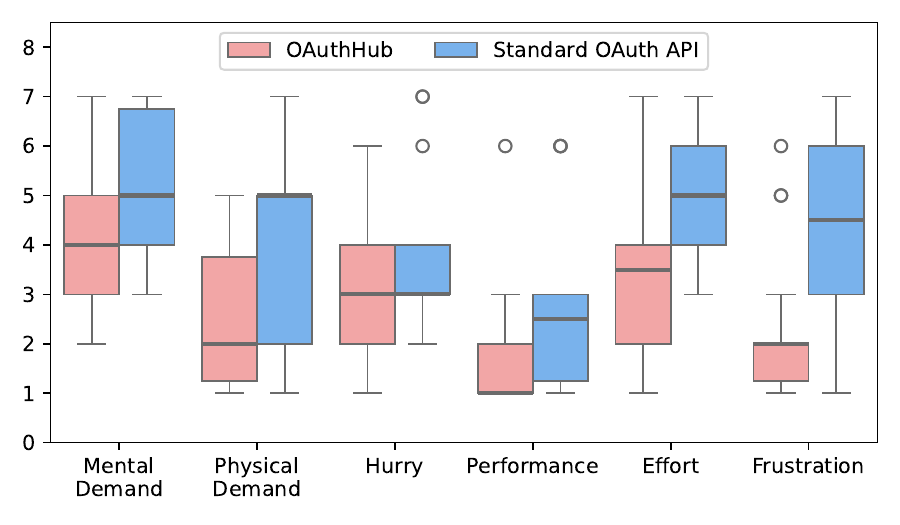}
    \vspace{-20pt}
    \caption{The cognitive load measured by NASA-TLX. Participants perceived significantly less mental load, effort and frustration using \sysname than conventional OAuth APIs.}
    \label{fig:nasa-tlx}
    % \vspace{-10pt}
\end{figure}

Additionally, we measured the perceived usability of APIs (Figure \ref{fig:nasa-tlx}). Participants perceived lower mental load, effort and frustration, along with better performance with \sysname than conventional OAuth APIs. These differences were statistically significant under a Wilcoxon Signed-Rank test (all p $<$ 0.01). 

\begin{table}[htbp]
    \centering\small
    \begin{tabular}{|c | c c | c c |}
    \hline
    \multirow{2}{*}{Task ID} & \multicolumn{2}{c|}{Conventional OAuth API} & \multicolumn{2}{c|}{\sysname} \\
    & Time & LOC & Time & LOC \\
    \hline 
    \hline
    Task 1 & 16.0 $\pm$ 5.8 & 19.2 $\pm$ 6.2 & \textbf{9.3 $\pm$ 3.6} & 5.3 $\pm$ 1.0 \\
    \hline
    Task 2 & 25.1 $\pm$ 10.9 & 13.6 $\pm$ 6.1 & \textbf{14.5 $\pm$ 5.5} & 6.7 $\pm$ 0.5 \\
    \hline
    Task 3 & 13.3 $\pm$ 4.7 & 8.2 $\pm$ 2.8 & \textbf{9.1 $\pm$ 2.8} & 5.9 $\pm$ 0.8 \\
    \hline
    Task 4 & 18.1 $\pm$ 3.8 & 24.6 $\pm$ 9.9 & \textbf{3.8 $\pm$ 1.8} & 1.6 $\pm$ 0.7 \\
    \hline
    
    \hline
    Average & 18.0 $\pm$ 5.4 & 15.8 $\pm$ 5.7 & \textbf{9.1 $\pm$ 3.0} & 4.7 $\pm$ 1.4 \\
    \hline
    \end{tabular}
    \caption{In the developer study, participants spent less time and used less code to complete the programming tasks with \sysname than conventional OAuth APIs. Format: mean $\pm$ standard deviation.
    % The results show that developers can complete programming tasks more easily and efficiently with \sysname.
    }
    \label{tab:developer_study}
    \vspace{-20pt}
\end{table}

In the post-study interview, participants provided more positive feedback on our API. When asked about their preferences between the two APIs, nearly all participants (16/18) leaned towards \sysname for programming, citing its ease of use over the original API provided by Google. The rest opted for conventional OAuth APIs because they were more familiar with the programming language or preferred procedural programming. However, they also envisioned that developers could complete programming tasks more efficiently with \sysname once getting familiar with its syntax, as it requires less code and is more concise. 

Participants noted several features of \sysname that contributed to their positive feedback. Most participants (12/18) appreciated the simple syntax of our manifest language, describing it as ``intuitive" and ``more readable" since each operator conveyed the high-level semantics of data transformation. Many participants (8/18) also appreciated its modular structure, likening it to SQL, and mentioned its advantages for analysis and debugging due to the clear separation of functions. 
Besides, some participants (6/18) valued the clarity of the documentation, finding it easy to navigate and helpful for grasping the language despite the initial learning curve. However, several participants (4/18) also noted the semantic limitations of our API language, which are intentionally designed to restrict data processing behaviors and mitigate potential vulnerabilities.

% As all participants had no prior knowledge of the GraphQL query language and we only gave them 10 minutes to learn our API language, we anticipate developers will gain familiarity with \sysname in real-world setting. 

\subsection{Mitigating User Privacy Concerns}
\label{sec:eval_user}
We evaluated whether \sysname can mitigate users' privacy concerns using an online survey based on three scenarios from our case studies (Section \ref{sec:case_study}).

\sssec{Method}. The survey used a between-subjects design where each participant experienced three scenarios in a randomized order. 
% We ensured that participants had prior experience using the app before each scenario; otherwise they could just skip it. 
For each scenario, we randomly presented each participant with one of the two interfaces.
% guided participants through the standard OAuth authorization process. 
We started by checking if they understood the implications of granting access to gauge their understanding of the permission requested by OAuth apps. 
Then, we asked how likely they would grant access to the OAuth app on a 5-point Likert scale. We included an open-ended question at the end of each scenario for users to explain the rationale behind their choices. The full survey questions are provided in Appendix~\ref{sec:survey}.

We recruited 100 participants through Cloud Research. We chose participants with an approval rate greater than 90\% who reside in the U.S. 
To ensure the quality of responses, we included an attention check at the end of the survey and filtered out 4 responses that did not pass. 
On average, the survey took each participant 4-6 minutes to complete.
Each participant received a compensation of 1 USD for their time. 
The study was approved by our institution's IRB.
% Our survey instructs users to perform several steps to experience the original OAuth page provided by data providers and third-party services. 
% Users are asked whether they will grant permissions given the scenario, as well as whether they understand the cases of data over-access in the given scenario. 
% After the explanation of the data over-access cases, users will be presented with the interactive interfaces of OAuthWall. They are then asked to choose if they will grant the permission to the same third-party service with the data over-access cases mitigated by OAuthWall. 
% The survey is designed using the Qualtrics platform and is distributed on Amazon MTurk in an anonymous manner.

\begin{figure}[htbp]
\includegraphics[width=\linewidth]{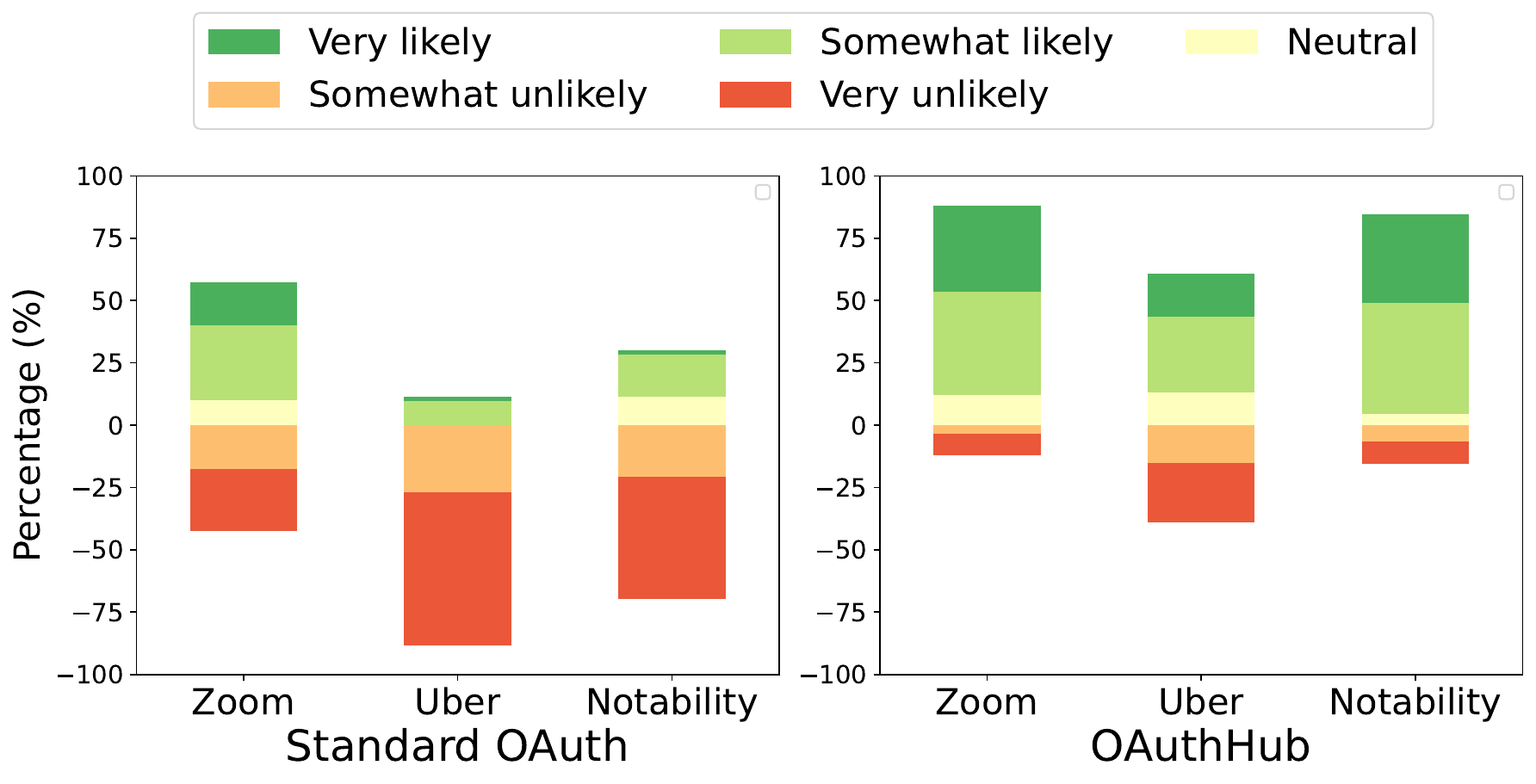}
\vspace{-20pt}
\caption{Users' likelihood of granting permissions for conventional OAuth requests and \sysname across three OAuth applications. Participants were more likely to approve the restricted permissions enabled by \sysname than conventional OAuth requests for all three OAuth applications.}
\label{fig:survey}
\vspace{-10pt}
\end{figure}

\sssec{Results}. Figure~\ref{fig:survey} presents the results of the survey. On average, 97\%, 94\% and 91\% of participants understood that conventional OAuth permissions would grant apps full access to personal data for Zoom, Uber and Notability, respectively. With \sysname, 97\%, 87\% and 96\% of participants correctly identified the restricted scope of permissions granted for these apps. The results suggest that users can clearly understand the permissions requested by OAuth apps to make informed decisions across different scenarios.

Over 40\% of users were more likely to deny the original OAuth permission requests for all three scenarios. We looked at open-ended responses to understand their decision rationale. Many participants were concerned that granting the permissions would allow the app to access more data than needed, especially when it involves sensitive information. For example, P16 noted that ``\textit{I don't think Uber needs access to all of my emails, some of which contain confidential material. I would allow them to have access to a separate folder that I send travel info to, but not all of my emails.}'' Participants also voiced concerns about the potential of modifying unrelated data: ``\textit{Given the fact that it can view and edit all of my calendar events, I would be worried about mistakes or errors it would make with my calendars. I would need to be able to view the changes that it makes in order to ensure that it is only focused on Zoom meetings and changes that are made in relation to that}'' (P62).

With \sysname, the likelihood of denial requests (i.e., somewhat or very unlikely to grant permissions) decreased from 42.5\% to 12.0\% in the Zoom scenario, with a reduction rate of 72\%. Similarly, the denial rates dropped by 56\% for Uber and 78\% for Notability. These results suggest that users are less likely to deny the fine-grained permission requests of \sysname than the original OAuth requests.

Participants' feedback also suggests that \sysname can help to alleviate users' privacy concerns. They appreciated the idea of limiting access to only the necessary data. P38 mentioned that ``\textit{It's not excessive and restricts itself to the links/contents that it should be concerned with, i.e., zoom links, not other info that it has no right to access.}"
Further, participants seemed to feel more comfortable with the fine-grained permissions enabled by \sysname: ``\textit{It is restricted to the content that is relevant, not all content which it is not entitled to access to use. If I needed to use the service / feature, then I would be more likely to proceed with this sort of limited access}" (P46) and 
``\textit{If I needed to use this app for a specific functionality / purpose, then I would feel comfortable being able to select a narrow scope of access versus allowing carte blanche access to everything}" (P55).

% We performed a Two Portion z-test and observed no statistical differences after accounting for multiple hypothesis correction. 

\subsection{System Performance}
\label{sec:eval_performance}
We used three example cases to quantify \sysname's system overhead in memory, CPU utilization and latency.
% , and energy consumption. 

\sssec{Method}. We tested all use cases on both PC and mobile settings.  For the PC, we used an Apple M1 MacBook Pro running Node.js v20.3.1. For mobile, we used a Huawei ELE-AL100 smartphone with 6GB memory, running Android 10.
For each use case, we requested the data from the client side of \sysname and measured the CPU utilization rates and peak memory usage during this process. 
We also measured the end-to-end latency from the moment the request was initiated until when the response was received. We executed each request 100 times to reduce potential inaccuracies caused by network fluctuations. We then compared the results with the latency when directly using the APIs provided by Google under the same test conditions. 

% while we directly use the google-api library package to perform the requests without proxying data through a local \sysname.

% We used the basic OAuth implementations provided by Google as the baseline, which handled requests directly with Google services.
% , and acquired the data from the mobile device via the app implementation. 

% performed the requests without proxying data through the local \sysname. 

For \textbf{energy consumption}, we charged the smartphone to full and connected it to a POWER-Z KT001 Portable USB-C Fast Charging Tester. We set the power plan of the smartphone to `performance first' to minimize measurement errors from system optimizations.
We then ensured the target application remained active for at least one hour, during which we recorded the average hourly energy usage. 
We conducted power consumption tests under four different conditions: 
(1) complete idle state, when all applications were shut down; (2) \sysname\ idle, when only \sysname was running but not executing any manifests; (3) \sysname Google Calendar $|$ Gmail $|$ Google Grive active, when \sysname was executing the manifest every 10 seconds; and (4) only a popular messenger app 
% (i.e., WeChat) 
was running in the background. 
All tests are done with screen on at 50\% brightness level.

\sssec{Results}. Table \ref{tab:overhead} shows the CPU and memory usage of \sysname under PC and mobile settings. \sysname consumed less than 1\% of CPU time and around 90MiB of memory on PC, and less than 5\% of CPU time with a peak memory usage of 200 MiB on the mobile device. Our results suggest that \sysname\ introduces modest CPU and memory overheads on the user's local device. 

% Given that our experiments involved a significantly higher frequency of operations compared to real-world scenarios constrained by API calling limits from the resource server, this level of overhead is deemed acceptable for modern local devices. 

\begin{figure}[htbp]
    \centering
    \begin{subfigure}[h]{0.5\linewidth}
    % \vspace{-5pt}
            \includegraphics[width=\textwidth]{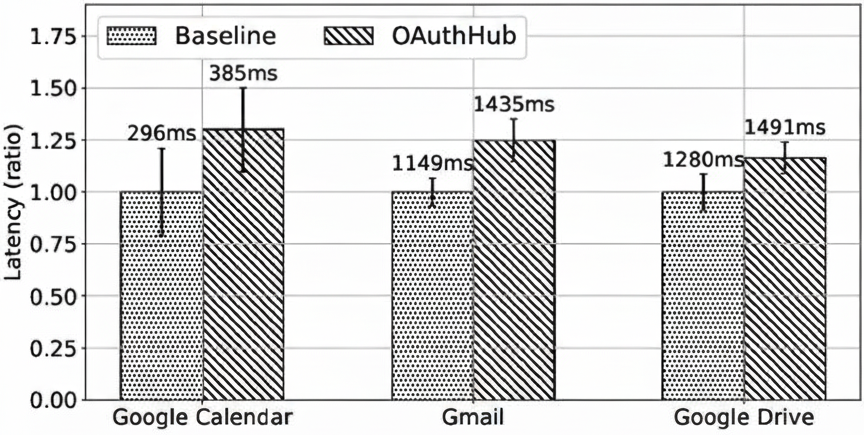}
            \caption{PC}
            \label{fig:latency_PC}
    \end{subfigure}
    \begin{subfigure}[h]{0.43\linewidth}
        \includegraphics[width=\textwidth]{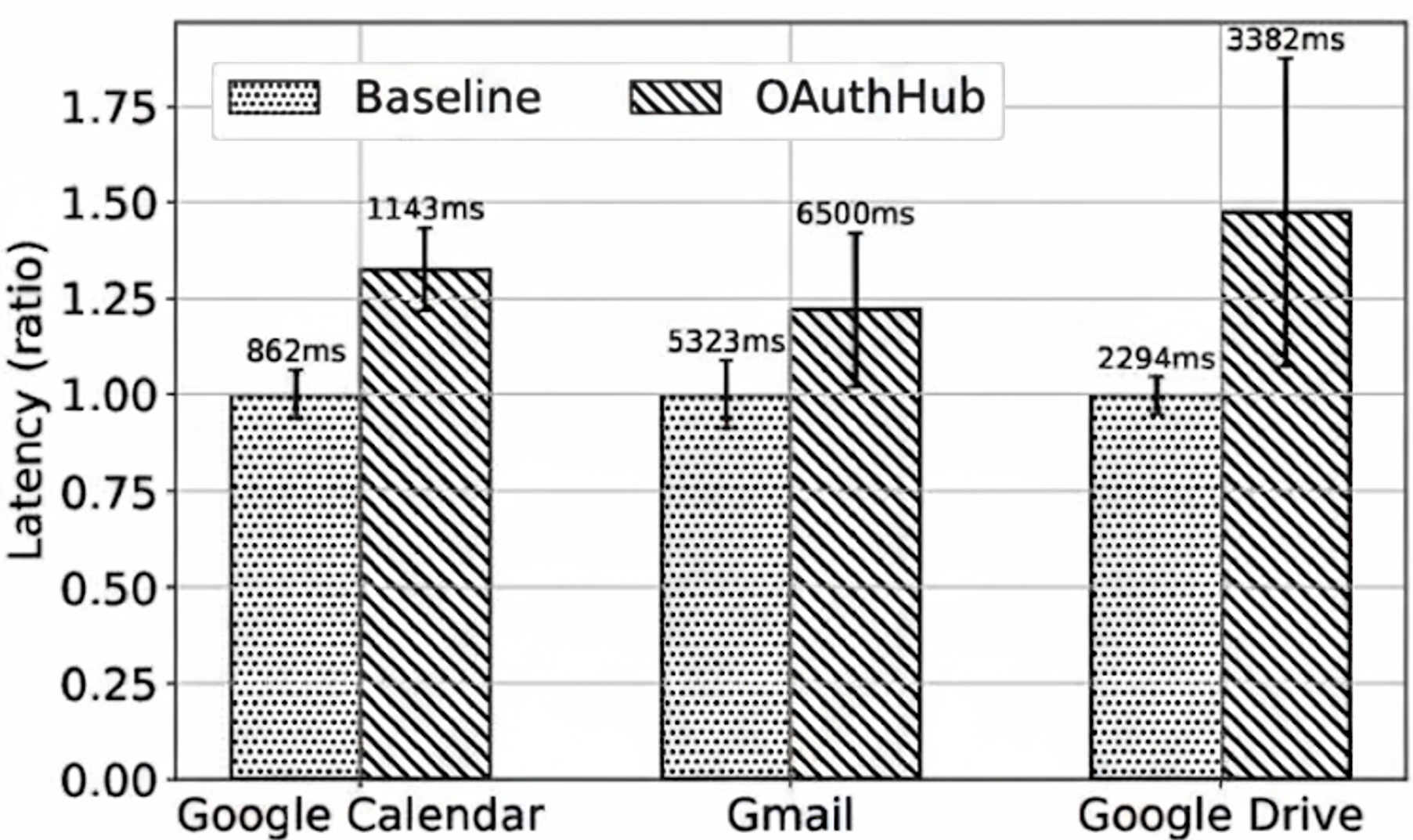}
        \caption{Mobile}
        \label{fig:latency_mobile}
    \end{subfigure}
    \vspace{-10pt}
   \caption{Average relative latency overhead of three OAuth applications on PC and mobile settings.
   % Absolute numbers for average latency in microseconds are annotated above each bar.
   }
  \label{fig:oauth-latency}  
\end{figure}

 Figure \ref{fig:oauth-latency} presents the latency of OAuth access with \sysname (dot bars) and without \sysname (slash bars). \sysname introduced relative overheads of 1.30, 1.25 and 1.16$\times$ on PC, and 1.33, 1.22 and 1.47$\times$ on mobile for Google Calendar, Gmail and Google Drive respectively. Google Drive exhibits the highest overhead (1.47$\times$) on mobile due to its resource-intensive nature, which might not be well-suited for mobile devices with lower processing power. We observed a lower relative overhead in Gmail on mobile because this app is more expensive as it handles a large volume of email messages, making the overhead less noticeable. On average, \sysname introduced an average latency of 195ms on PC and 849ms on the mobile device.

 % \sysname introduces around a 50 to 200ms latency of overhead for each request compared to directly pulling data from the resource server. 

 % Considering the frequency in our benchmark is much higher than real scenarios with the API calling limits from the resource server, the overhead is acceptable in modern local devices.
 
 % Our experiments indicate that \sysname\ introduces modest performance overheads on user's local device. 

% We chose three end-to-end use cases (Google Calendar, Gmail and Google Drive) to evaluate the system's latency. We chose these end-to-end tasks because they represent different types of pre-processing (i.e., filtering fields, multiple data pulling, and file write restriction). 

% All three end-to-end use cases were tested with Node.js v18.17.1 (libnodejs v18.17.1 wrapped in a minimal application in mobile settings). 
% The desktop data is collected from an i7-9700 desktop PC with 64GB memory, and 
% the mobile data is collected from a  The AMQP service for reverse proxy is provided by CloudAmqp with Amazon Web Services. 

% \noindent\textit{Energy}: 

The energy overhead results are shown in Table \ref{tab:energy}. \sysname causes a 12\% to 39\% energy overhead when running on a mobile phone, depending on the specific workload. The seemingly high percentages are mostly a result of our run-nothing blank control group. Actually, in the same testing environment, this overhead is similar to the one caused by running a messenger app.
% Nowadays, mobile phones rarely have nothing to run in the background, and therefore, in a real use case with several background tasks, the energy overhead of \name is acceptable or even negligible.

\begin{table}[htbp]
\centering\small
    \begin{tabular}{|c|c c|c c|}
    \hline
    Scenario  & \multicolumn{2}{c|}{PC} &  \multicolumn{2}{c|}{Mobile}\\
    & CPU & Memory & CPU & Memory \\
    \hline
    \hline
    Google Calendar  & $<1\%$ &  90.6 MiB & $<1\%$ & 181.5 MiB \\
             
    \hline         
    Gmail  & $<1\%$ & 92.5 MiB & $<5\%$ & 197.3 MiB\\
    \hline
    Google Drive & $<1\%$ & 91.6 MiB & $<3\%$ & 188.6 MiB\\ 
    \hline
    \end{tabular}
    \caption{Performance overhead of three use cases using \sysname when responding to 100 consecutive requests. CPU utilization rate represents the average percentage of active time consumed for 100 consecutive requests and memory indicates the peak memory usage within this process.}
    \label{tab:overhead}
    \vspace{-20pt}
\end{table}

% The latency data are collected by averaging 100 repeated requests from our emulated third-party service to users' devices. The standard error is illustrated with error bars. Each pair of data from three use cases are significantly different using Mann–Whitney U test ($p < 0.001$)

% \subsection{System energy consumption overhead}

\begin{table}[h!]
\centering\small
    \begin{tabular}{|c|c|c|c|}
    \hline
    Benchmark & Energy & $\Delta$ energy &  Overhead\\
    Scenario & (Joule/Hour) & (Joule/Hour) & (\%)\\
    \hline
    \hline
    Nothing running & 3312 & - & - \\
    \hline
    \sysname idle & 3708 & 396 & 11.96\\

    \hline
    Google Calendar active & 3816 & 504& 15.23\\
    Gmail active & 4608 & 1296 & 39.13\\
    Google Drive active & 3960 & 648 & 19.56\\
    \hline
    A messenger app running & 3852 & 540 & 16.30 \\
    \hline
    \end{tabular}
    \caption{Energy consumption overhead of three use cases using \sysname.
    % (1) Energy consumption is the raw energy consumption data. 
    % (2) $\Delta$ energy consumption is the difference of joule per hour in each scenario compared with the Nothing running scenario. 
    % (3) Energy consumption overhead is the percentage of extra energy consumed in an hour compared with the Nothing running scenario.
    % (4) Nothing running scenario is the baseline energy consumption where no app is running. 
    % (5) OAuthWall idle is the case where OAuthWall is running but handles no requests.
    % (6) OAuthWall active cases are those where OAuthWall is handling 1 data request per 10 seconds. 
    % (7) WeChat running case is where only WeChat is running.
    }
    \label{tab:energy}
    \vspace{-20pt}
\end{table}

\section{Related Work}\label{sec:related}

\sssec{Data Overaccess in Commericial Protocols}. 
Many studies have studied the security and privacy of commercial protocols~\cite{sun2012devil, yang2013security,sadqi2020web,fett2016comprehensive,al2019oauthlint, li2019oauthguard, singh2022oauth, chen2014oauth, corre2017can, wijayarathna2019empirical}. 
The most relevant line of work is to study the overaccess problems in OAuth and IFTTT~\cite{chen2022practical}, where app developers request data access that they do not necessarily need~\cite{morkonda2021empirical, dimova2023everybody,li2020user,fernandes2018decentralized,jin2022peekaboo,cao2024stateful,chen2022practical}. For example, Chen et al.~\cite{chen2022practical} identified two key design flaws in IFTTT that can cause data privacy problems: attribute level overprivilege and token-level overprivilege. 
In contrast to these works, \sysname\ focuses on the temporal domain and demonstrates that discretionary access to user data through OAuth is an avoidable privilege. 

Several works have specifically analyzed privacy issues in OAuth and proposed tools to detect identity leaks~\cite{jannett2024sok, sun2012devil}. For example, SSO‑Monitor provides large‑scale monitoring of Single Sign‑On implementations and their associated OAuth flows, collecting metrics on security and privacy practices across millions of websites to identify insecure or improper configurations~\cite{jannett2024sok}. In contrast to existing tools that monitor and analyze identity information leaks, our work enables developers to proactively limit data access.

\sssec{Manifest \& Operator-based API}.
Many studies use manifests to declare permissions and to create a sandboxed environment to contain untrusted applications (e.g., Janus~\cite{goldberg1996secure}, Android permissions~\cite{Networks78:online}, MUDs~\cite{WhatisMU82:online}). 
Conventional manifests often confine applications' binary access to system resources. 
For example, Android developers may declare the need for all-or-nothing access to users' contacts.

However, due to the vast number of data granularities and the nuanced contextual nature of privacy, this traditional attribute-based manifest representation is becoming increasingly insufficient. Researchers propose a few approaches to enhance the permission system. One relevant approach is through remote code execution~\cite{de2014openpds,cao2024stateful}. For example, Cao et al.~\cite{cao2024stateful} proposed allowing app developers to author data-filtering programs, which are then executed by service providers through in-process sandboxing. 
While the remote code execution offers great flexibility for adding new data accesses, it is hard to enforce (i.e., analyze and control) the data transformation behaviors of these arbitrary programs. 
In contrast, \sysname\ only allows developers to specify data collection behaviors using a set of high-level operators with well-defined semantics. So \sysname\ can easily infer the program behaviors by analyzing the manifest and enabling a centralized runtime permission system.

\sssec{Local Hub}.
Centralizing security and privacy management through a local hub is a widely adoptedapproach~\cite{zhang2021capture,jin2022peekaboo,zavalyshyn2022sok}. For instance, Capture~\cite{zhang2021capture} uses a centralized hub to manage and deploy IoT device firmware. The hub design of \sysname\ is particularly inspired by Peekaboo~\cite{jin2022peekaboo}, sharing several key elements, such as operator-based manifest design and centralized end-user management features. However, \sysname\ introduces several key differences. Unlike Peekaboo, which operates as an always-on hub that only transmits data and does not receive callbacks from servers, \sysname\ is designed to run on intermittently available personal devices and supports both read and write access.
Further, \sysname\ explores a new runtime permission mechanism for OAuth apps.

\section{Discussion}
\label{sec:discussion}

\sssec{Benefits of \sysname}. 
Having a local hub mediating the OAuth-based sharing has three advantages over the conventional OAuth implementation. 
First, \sysname enables fine-grained data access, even if the service providers do not offer such. 
Second, users have better control and transparency about how OAuth apps access their data. 
Third, \sysname helps developers reflect when their apps need to access user data. 

\sssec{Alternatives}. We also consider alternative solutions to using the user’s device as a local data hub, such as cloud-based approaches and user-managed serverless runtimes. For example, server-side filtering in a trusted execution environment prevents third parties from accessing raw data, which offers better security but increases computational overhead. A user-owned serverless runtime improves availability but requires technical expertise for setup.

The data minimization enabled by \sysname is complementary to these cloud-based privacy solutions. OAuth has become popular due to its lightweight design. Most OAuth developers may not have the expertise to utilize advanced privacy-enhancing technologies such as Google Confidential Computing~\cite{googleconfidential:Online}.

\sssec{\sysname adoption}. In high-stakes domains (e.g., applications handling sensitive personal data), developers have strong incentives to earn and maintain user trust, since overly aggressive permissions can trigger denials and damage an app’s reputation~\cite{tahaei2023stuck}.

\sysname has three key ideas: (1) a data hub that mediates OAuth-based data sharing, (2) using users' personal devices as the hub, and (3) operator-based data access. 
We expect that the biggest adoption barrier of \sysname is requiring users to install the hub runtime (e.g., a Chrome extension or a mobile app) on their devices if \sysname is not a built-in service.

A promising approach to address this challenge is establishing a serverless data hub operating on a personal cloud~\cite{palkar2017diy,kim2022self}. However, this solution introduces a privacy tradeoff: the service provider (e.g., AWS) will know what OAuth services the user has interacted with. 
An alternative is self-hosting the data hub. Self-hosting is increasingly popular among privacy-conscious users (e.g., Vaultwarden~\cite{vaultwarden:Online}), as it avoids reliance on third-party cloud providers and gives users full control over their data. It can also provide additional benefits such as always-available OAuth processing, reduced latency within their local network, and centralized management across multiple personal devices. At the same time, self-hosting requires technical expertise and ongoing maintenance, which may limit its accessibility to non-expert users.

\sssec{Integration of \sysname}. To integrate \sysname, developers add a new ``Sign in with OAuthHub'' option and register an API endpoint for receiving tokens and data from the local hub, with rate limiting and robust DDoS protection. The integration mainly relies on the three library APIs shown in Figure~\ref{fig:lib}: (1) \texttt{generateAuthUrl} to initiate authorization with a fine-grained manifest and declared access pattern, (2) \texttt{exchangeToken} to receive the access token via the app’s callback endpoint, and (3) \texttt{query} to retrieve or write data through \sysname instead of directly calling service-provider APIs. 
Developers also implement server-side logic to verify payloads from the local hub and include error handling for offline devices.
Appendix~\ref{sec:code_changes} illustrates code changes for the Zoom scenario.

% \qiyu{error handling, server-side logic, and UI modifications}

\sssec{Decentralized web}. Today, service providers like Google manage users' data on their servers. This paradigm minimizes the burden for users but introduces a few key privacy concerns. More recently, web pioneers are advocating the idea of Web5, where users' devices (nodes) store and process data locally, with the server only caching data for service availability when the local device is unavailable~\cite{web5}. 
However, most users do not have the expertise to maintain the availability of their nodes, and most Web5 services are incompatible with existing services. 
\sysname represents a balanced approach to the Web5 vision, offering a practical middle ground where the local hub delivers a substantial level of privacy protection and user control while remaining feasible for implementation.

\sssec{Cross-device integration}.
The current implementation of \sysname\ operates local runtimes independently on individual devices, typically a PC and a smartphone for most users. 
Users have to grant access to the same service separately across devices, the same as the current OAuth implementation. 
A potential advantage of the previously discussed personal cloud version of \sysname\ is its ability to unify management across multiple devices, where users only need to authorize each app once across devices.

\sssec{Service reliability}.
% \sysname operates per device, managing only the OAuth data flows on that device. 
Each device runs its own \sysname instance if the user logs in through \sysname on multiple devices. Losing a phone does not affect service on other devices. 
The insight behind \sysname is that most OAuth apps can wait to fetch information until users actually need to use the device, when the device becomes online. Even after a prolonged offline period, the device will fetch data when the user opens the service on that device. For certain special cases that involve time-sensitive scheduled tasks across devices, such as when a user schedules a Drive backup on their phone but later needs to access the file from their laptop while the phone is offline, \sysname does not support this scenario.
% However, \sysname does not support scheduled tasks that involve time-sensitive operations across devices, e.g., a user schedules a Drive backup on their phone but needs to access the file from their laptop while the phone is offline.

\vspace{-6pt}

\section{Limitations}\label{sec:limitation}

\sysname is not designed to support \textbf{data-intensive OAuth access}. For instance, transferring multi-gigabyte files through the local data hub can be slow and is better handled via conventional OAuth protocols. Similarly, an app like draw.io, which may want to save drawing files to Google Drive after each minor edit, may lead to significant power consumption for the personal device. Fortunately, our empirical analysis of real-world OAuth applications has not revealed these data-intensive OAuth usages.

\noindent\textbf{Data schema}. Service providers often implement diverse data schemas, requiring manual efforts to define and unify them within \sysname. Despite these differences, our findings indicate that most providers share common semantic structures. However, achieving full interoperability might be a significant engineering challenge. Future work could explore leveraging generative AI techniques to automate schema unification~\cite{tian2023chatgpt}.

\sssec{Maintenance efforts}. \sysname supports updates from third-party applications and service providers. To change the use case or requested data (e.g., adding new providers or supporting multiple languages), developers can submit a new manifest for new providers or update the existing manifest (e.g., adding keywords). Any manifest update must go through the authorization process again. Additionally, if OAuth endpoints change their data schema, \sysname needs to update the code that maps input data to the predefined schema. 
We expect \sysname to require only periodic updates rather than frequent fixes for breaking API changes, as major OAuth service providers maintain backward compatibility with long deprecation periods to minimize disruptions.

% Breaking updates are rare because major OAuth providers maintain backward compatibility with long deprecation periods. For example, Google deprecated the Drive Android API in 2018, stopped new usage in 2019, and did not affect existing integrations until its full shutdown in 2023. Because OAuth services are incentivized to minimize disruptions, we expect OAuthHub only make periodic updates rather than frequent fixes for breaking API changes.

\sssec{Measuring user preferences}. 
We used surveys to understand how likely users will deny an OAuth request. However, these surveys may introduce biases, as previous research found that survey participants often align their answers with social expectations to some extent~\cite{kokolakis2017privacy,alsoubai2022permission}.Despite these limitations, survey responses still provide meaningful insights into user expectations~\cite{redmiles2019well}, particularly their desire for more fine-grained data access controls.

However, the expected behavior studied may differ from real-world use. In real-world scenarios, users may act differently, either due to a lack of attention to privacy controls or because they perceive the benefits of a service outweigh its potential risks. We plan to monitor the open-source adoption, and interview users in the wild to understand their perspectives in future work.

\sssec{Sampling bias}. The OAuth apps evaluated were primarily sourced from workplace-oriented marketplaces, which may be biased towards corporate environments. We plan to extend our evaluation to additional domains in future work (e.g., social networks).

\vspace{-10pt}

% \qiyu{While the marketplace includes many business-oriented apps, it also includes apps that handle personal data (e.g., WhatsApp Messenger). acknowledge the potential sampling bias and plan to extend the evaluation to additional domains (e.g., social networking).}

% Our evaluation of mitigating user privacy concerns might be biased. Previous research found that users are more likely to behave according to society's expectations~\cite{}. So this may exaggerate the deny rate comparing to the deny rate in the real world. 

% [Be careful about the argument. Do not write the section like our study is useless.]

\section{Conclusion}
\label{sec:conclusion}
This paper introduces \sysname, a development framework that uses personal devices as the intermediary controller for data sharing between cloud services. A key challenge is that these devices are not always online and have no static IP.  To address this, \sysname leverages a new insight that most apps need data access only at install time, user-driven events, or scheduled times. \sysname requires developers to declare when they will request data explicitly, and mediates these data access using not-always-on local devices. The introduction of local data hubs enables a centralized runtime permission model for managing OAuth access across services. 
We developed \sysname as an open-source library for OAuth app developers to integrate.
Our evaluation with real-world apps shows that \sysname requires minimal changes to the application code, and is easy for developers to use compared to the conventional OAuth APIs.

\bibliographystyle{ACM-Reference-Format}
\bibliography{sample-base}

@techreport{rfcoauth1.0,
  author = {Hammer-Lahav, Eran},
  title = {The OAuth 1.0 Protocol},
  howpublished = {Internet Requests for Comments},
  type = {RFC},
  number = 5849,
  year = {2010},
  month = {04},
  issn = {2070-1721},
  publisher = {RFC Editor},
  institution = {RFC Editor},
  url = {https://www.rfc-editor.org/rfc/rfc5849}
}

@techreport{rfcoauth2.0,
  author = {Hardt, Dick},
  title = {The OAuth 2.0 Authorization Framework},
  howpublished = {Internet Requests for Comments},
  type = {RFC},
  number = 6749,
  year = {2012},
  month = {10},
  issn = {2070-1721},
  publisher = {RFC Editor},
  institution = {RFC Editor},
  url = {https://www.rfc-editor.org/rfc/rfc6749}
}

@techreport{rfcoauthnative,
  author = {W. Denniss and Google and J. Bradley and Ping Identity},
  title = {OAuth 2.0 for Native Apps},
  howpublished = {Internet Requests for Comments},
  type = {RFC},
  number = 8252,
  year = {2017},
  month = {10},
  issn = {2070-1721},
  publisher = {RFC Editor},
  institution = {RFC Editor},
  url = {https://datatracker.ietf.org/doc/html/rfc8252}
}

@techreport{rfcpkce,
  author = {N. Sakimura, Ed. and Nomura Research Institute and J. Bradley and Ping Identity and N. Agarwal and Google},
  title = {Proof Key for Code Exchange by OAuth Public Clients},
  howpublished = {Internet Requests for Comments},
  type = {RFC},
  number = 7636,
  year = {2015},
  month = {09},
  issn = {2070-1721},
  publisher = {RFC Editor},
  institution = {RFC Editor},
  url = {https://www.rfc-editor.org/rfc/rfc7636}
}

@article{leiba2012oauth,
  title={Oauth web authorization protocol},
  author={Leiba, Barry},
  journal={IEEE Internet Computing},
  volume={16},
  number={1},
  pages={74--77},
  year={2012},
  publisher={IEEE}
}

@inproceedings{philipps99,
  title={Refinement of pipe-and-filter architectures},
  author={Philipps, Jan and Rumpe, Bernhard},
  booktitle={FM’99—Formal Methods: World Congress on Formal Methods in the Development of Computing Systems Toulouse, France, September 20--24, 1999 Proceedings, Volume I},
  pages={96--115},
  year={1999},
  organization={Springer}
}

@techreport{RFC9396O83:RichAuth,
  author = {T. Lodderstedt and J. Richer and B. Campbell
},
  title = {RFC 9396 - OAuth 2.0 Rich Authorization Requests},
  howpublished = {Internet Requests for Comments},
  type = {RFC},
  number = 5849,
  year = {2023},
  month = {05},
  publisher = {RFC Editor},
  institution = {RFC Editor},
  url = {https://datatracker.ietf.org/doc/html/rfc9396}
}

@inproceedings{fett2016comprehensive,
  title={A comprehensive formal security analysis of OAuth 2.0},
  author={Fett, Daniel and K{\"u}sters, Ralf and Schmitz, Guido},
  booktitle={Proceedings of the 2016 ACM SIGSAC Conference on Computer and Communications Security},
  pages={1204--1215},
  year={2016}
}

@inproceedings{yang2013security,
  title={A security analysis of the OAuth protocol},
  author={Yang, Feng and Manoharan, Sathiamoorthy},
  booktitle={2013 IEEE Pacific Rim Conference on Communications, Computers and Signal Processing (PACRIM)},
  pages={271--276},
  year={2013},
  organization={IEEE}
}

@inproceedings{sun2012devil,
  title={The devil is in the (implementation) details: an empirical analysis of OAuth SSO systems},
  author={Sun, San-Tsai and Beznosov, Konstantin},
  booktitle={Proceedings of the 2012 ACM conference on Computer and communications security},
  pages={378--390},
  year={2012}
}

@inproceedings{al2019oauthlint,
  title={Oauthlint: An empirical study on oauth bugs in android applications},
  author={Al Rahat, Tamjid and Feng, Yu and Tian, Yuan},
  booktitle={2019 34th IEEE/ACM International Conference on Automated Software Engineering (ASE)},
  pages={293--304},
  year={2019},
  organization={IEEE}
}

@inproceedings{li2020user,
  title={User access privacy in OAuth 2.0 and OpenID connect},
  author={Li, Wanpeng and Mitchell, Chris J},
  booktitle={2020 IEEE European Symposium on Security and Privacy Workshops (EuroS\&PW)},
  pages={664--6732},
  year={2020},
  organization={IEEE}
}

@inproceedings{jin2022peekaboo,
  title={Peekaboo: A hub-based approach to enable transparency in data processing within smart homes},
  author={Jin, Haojian and Liu, Gram and Hwang, David and Kumar, Swarun and Agarwal, Yuvraj and Hong, Jason I},
  booktitle={2022 IEEE Symposium on Security and Privacy (SP)},
  pages={303--320},
  year={2022},
  organization={IEEE}
}

@misc{WhatisMU82:online,
author = {Cisco},
title = {What is MUD? - Manufacturer Usage Description - Document - Cisco DevNet},
howpublished = {\url{https://developer.cisco.com/docs/mud/##!what-is-mud}},
month = {02},
year = {2021},
note = {(Accessed on 02/05/2021)}
}

@misc{Networks78:online,
author = {Android Developer},
title = {Network security configuration  |  Android Developers},
howpublished = {\url{https://developer.android.com/training/articles/security-config}},
month = {07},
year = {2020},
note = {(Accessed on 09/07/2020)}
}

@inproceedings{roesner2012user,
  title={User-driven access control: Rethinking permission granting in modern operating systems},
  author={Roesner, Franziska and Kohno, Tadayoshi and Moshchuk, Alexander and Parno, Bryan and Wang, Helen J and Cowan, Crispin},
  booktitle={2012 IEEE Symposium on Security and Privacy},
  pages={224--238},
  year={2012},
  organization={IEEE}
}

@inproceedings{goldberg1996secure,
  title={A secure environment for untrusted helper applications: Confining the wily hacker},
  author={Goldberg, Ian and Wagner, David and Thomas, Randi and Brewer, Eric A and others},
  booktitle={Proceedings of the 6th conference on USENIX Security Symposium, Focusing on Applications of Cryptography},
  volume={6},
  pages={1--1},
  year={1996}
}

@inproceedings{chen2014oauth,
  title={Oauth demystified for mobile application developers},
  author={Chen, Eric Y and Pei, Yutong and Chen, Shuo and Tian, Yuan and Kotcher, Robert and Tague, Patrick},
  booktitle={Proceedings of the 2014 ACM SIGSAC conference on computer and communications security},
  pages={892--903},
  year={2014}
}

@article{privacystreams,
author = {Li, Yuanchun and Chen, Fanglin and Li, Toby Jia-Jun and Guo, Yao and Huang, Gang and Fredrikson, Matthew and Agarwal, Yuvraj and Hong, Jason I.},
title = {PrivacyStreams: Enabling Transparency in Personal Data Processing for Mobile Apps},
year = {2017},
issue_date = {September 2017},
publisher = {Association for Computing Machinery},
address = {New York, NY, USA},
volume = {1},
number = {3},
url = {https://doi.org/10.1145/3130941},
doi = {10.1145/3130941},
journal = {Proc. ACM Interact. Mob. Wearable Ubiquitous Technol.},
month = sep,
articleno = {76},
numpages = {26}
}

@inproceedings{sadqi2020web,
  title={Web oauth-based sso systems security},
  author={Sadqi, Yassine and Belfaik, Yousra and Safi, Said},
  booktitle={Proceedings of the 3rd International Conference on Networking, Information Systems \& Security},
  pages={1--7},
  year={2020}
}

@inproceedings{li2019oauthguard,
  title={Oauthguard: Protecting user security and privacy with oauth 2.0 and openid connect},
  author={Li, Wanpeng and Mitchell, Chris J and Chen, Thomas},
  booktitle={Proceedings of the 5th ACM workshop on security standardisation research workshop},
  pages={35--44},
  year={2019}
}

@article{singh2022oauth,
  title={OAuth 2.0: Architectural design augmentation for mitigation of common security vulnerabilities},
  author={Singh, Jaimandeep and Chaudhary, Naveen Kumar},
  journal={Journal of Information Security and Applications},
  volume={65},
  pages={103091},
  year={2022},
  publisher={Elsevier}
}

@inproceedings{wijayarathna2019empirical,
  title={An empirical usability analysis of the google authentication api},
  author={Wijayarathna, Chamila and Arachchilage, Nalin AG},
  booktitle={Proceedings of the 23rd International Conference on Evaluation and Assessment in Software Engineering},
  pages={268--274},
  year={2019}
}

@misc{calendar:online,
author = {Google},
title = {Choose Google Calendar API scopes},
howpublished = {\url{https://developers.google.com/calendar/api/auth?hl=en}},
month = {09},
year = {2023},
note = {(Accessed on 14/09/2023)}
}

@inproceedings{jain2014should,
  title={Should I protect you? Understanding developers’ behavior to privacy-preserving APIs},
  author={Jain, Shubham and Lindqvist, Janne and others},
  booktitle={Workshop on Usable Security (USEC’14)},
  year={2014}
}

@misc{Howtoacc61:online,
author = {{Citizens Information}},
title = {How to access your personal data under the GDPR},
howpublished = {\url{https://www.citizensinformation.ie/en/government-in-ireland/data-protection/rights-under-general-data-protection-regulation/}},
month = {11},
year = {2023},
note = {(Accessed on 11/30/2023)}
}

@misc{UsingOAu5:online,
author = {Google},
title = {Using OAuth 2.0 to Access Google APIs},
howpublished = {\url{https://developers.google.com/identity/protocols/oauth2}},
month = {11},
year = {2023},
note = {(Accessed on 11/30/2023)}
}

@inproceedings{balash2022security,
  title={Security and Privacy Perceptions of $\{$Third-Party$\}$ Application Access for Google Accounts},
  author={Balash, David G and Wu, Xiaoyuan and Grant, Miles and Reyes, Irwin and Aviv, Adam J},
  booktitle={31st USENIX Security Symposium (USENIX Security 22)},
  pages={3397--3414},
  year={2022}
}

@inproceedings{bloch2006design,
  title={How to design a good API and why it matters},
  author={Bloch, Joshua},
  booktitle={Companion to the 21st ACM SIGPLAN symposium on Object-oriented programming systems, languages, and applications},
  pages={506--507},
  year={2006}
}

@inproceedings{hart2006nasa,
  title={NASA-task load index (NASA-TLX); 20 years later},
  author={Hart, Sandra G},
  booktitle={Proceedings of the human factors and ergonomics society annual meeting},
  volume={50},
  number={9},
  pages={904--908},
  year={2006},
  organization={Sage publications Sage CA: Los Angeles, CA}
}

@misc{appauth:online,
author = {AppAuth},
title = {AppAuth--Native App SDK for OAuth 2.0 and OpenID Connect implementing modern best practices},
year={2021},
month={12},
howpublished = {\url{https://appauth.io/}},
note = {(Accessed on 09/04/2025)}
}

@misc{google:online,
author = {Google},
title = {Using OAuth 2.0 to Access Google APIs},
howpublished = {\url{https://developers.google.com/identity/protocols/oauth2\#expiration
}},
month = {05},
year = {2024},
note = {(Accessed on 06/04/2024)}
}

@inproceedings{morkonda2021empirical,
  title={Empirical analysis and privacy implications in OAuth-based single sign-on systems},
  author={Morkonda, Srivathsan G and Chiasson, Sonia and van Oorschot, Paul C},
  booktitle={Proceedings of the 20th Workshop on Workshop on Privacy in the Electronic Society},
  pages={195--208},
  year={2021}
}

@article{dimova2023everybody,
  title={Everybody's Looking for SSOmething: A large-scale evaluation on the privacy of OAuth authentication on the web},
  author={Dimova, Yana and Van Goethem, Tom and Joosen, Wouter},
  journal={Proceedings on Privacy Enhancing Technologies},
  year={2023}
}

@misc{community_hubspot_2022, url={https://community.hubspot.com/t5/APIs-Integrations/Does-the-OAuth-refresh-token-expire/m-p/335543}, journal={community.hubspot.com}, year={2022}, month={May}}

@misc{androidManifestOverview,
	title = {{A}pp manifest overview  |  {A}ndroid {D}evelopers --- developer.android.com},
	howpublished = {\url{https://developer.android.com/guide/topics/manifest/manifest-intro}},
	year = {2024},
	note = {[Accessed 06-06-2024]},
}

@inproceedings{hartig2018semantics,
  title={Semantics and complexity of GraphQL},
  author={Hartig, Olaf and P{\'e}rez, Jorge},
  booktitle={Proceedings of the 2018 World Wide Web Conference},
  pages={1155--1164},
  year={2018}
}

@inproceedings{fokaefs2011empirical,
  title={An empirical study on web service evolution},
  author={Fokaefs, Marios and Mikhaiel, Rimon and Tsantalis, Nikolaos and Stroulia, Eleni and Lau, Alex},
  booktitle={2011 IEEE International Conference on Web Services},
  pages={49--56},
  year={2011},
  organization={IEEE}
}

@inproceedings{shen2021can,
  title={Can systems explain permissions better? understanding users' misperceptions under smartphone runtime permission model},
  author={Shen, Bingyu and Wei, Lili and Xiang, Chengcheng and Wu, Yudong and Shen, Mingyao and Zhou, Yuanyuan and Jin, Xinxin},
  booktitle={30th USENIX Security Symposium (USENIX Security 21)},
  pages={751--768},
  year={2021}
}

@misc{web5,
author = {Angie Jones},
title = {What is Web5?},
howpublished = {\url{https://dev.to/tbdevs/what-is-web5-233o}},
month = {07},
year = {2022},
note = {(Accessed on 11/01/2024)}
}

@article{hu2015attribute,
  title={Attribute-based access control},
  author={Hu, Vincent C and Kuhn, D Richard and Ferraiolo, David F and Voas, Jeffrey},
  journal={Computer},
  volume={48},
  number={2},
  pages={85--88},
  year={2015},
  publisher={IEEE}
}

@inproceedings{felt2011effectiveness,
  title={The effectiveness of application permissions},
  author={Felt, Adrienne Porter and Greenwood, Kate and Wagner, David},
  booktitle={2nd USENIX Conference on Web Application Development (WebApps 11)},
  year={2011}
}

@inproceedings{felt2011android,
  title={Android permissions demystified},
  author={Felt, Adrienne Porter and Chin, Erika and Hanna, Steve and Song, Dawn and Wagner, David},
  booktitle={Proceedings of the 18th ACM conference on Computer and communications security},
  pages={627--638},
  year={2011}
}

@inproceedings{tian2017smartauth,
  title={$\{$SmartAuth$\}$:$\{$User-Centered$\}$ authorization for the internet of things},
  author={Tian, Yuan and Zhang, Nan and Lin, Yueh-Hsun and Wang, XiaoFeng and Ur, Blase and Guo, Xianzheng and Tague, Patrick},
  booktitle={26th USENIX Security Symposium (USENIX Security 17)},
  pages={361--378},
  year={2017}
}

@misc{KeyStore:online,
author = {Android },
title = {Android Keystore system},
howpublished = {\url{https://developer.android.com/privacy-and-security/keystore}},
month = {09},
year = {2024},
note = {(Accessed on 01/06/2024)}
}

@inproceedings{andriotis2016permissions,
  title={Permissions snapshots: Assessing users' adaptation to the Android runtime permission model},
  author={Andriotis, Panagiotis and Sasse, Martina Angela and Stringhini, Gianluca},
  booktitle={2016 IEEE International Workshop on Information Forensics and Security (WIFS)},
  pages={1--6},
  year={2016},
  organization={IEEE}
}

@inproceedings{nauman2010apex,
  title={Apex: extending android permission model and enforcement with user-defined runtime constraints},
  author={Nauman, Mohammad and Khan, Sohail and Zhang, Xinwen},
  booktitle={Proceedings of the 5th ACM symposium on information, computer and communications security},
  pages={328--332},
  year={2010}
}

@inproceedings{wei2012permission,
  title={Permission evolution in the android ecosystem},
  author={Wei, Xuetao and Gomez, Lorenzo and Neamtiu, Iulian and Faloutsos, Michalis},
  booktitle={Proceedings of the 28th Annual Computer Security Applications Conference},
  pages={31--40},
  year={2012}
}

@inproceedings{peruma2018investigating,
  title={Investigating user perception and comprehension of android permission models},
  author={Peruma, Anthony and Palmerino, Jeffrey and Krutz, Daniel E},
  booktitle={Proceedings of the 5th International Conference on Mobile Software Engineering and Systems},
  pages={56--66},
  year={2018}
}

@misc{Jenkins:online,
title = {Jenkins plugins index},
howpublished = {\url{https://plugins.jenkins.io/}},
year={2025},
note = {(Accessed on 01/20/2025)}
}

@inproceedings{palkar2017diy,
  title={Diy hosting for online privacy},
  author={Palkar, Shoumik and Zaharia, Matei},
  booktitle={Proceedings of the 16th ACM Workshop on Hot Topics in Networks},
  pages={1--7},
  year={2017}
}

@article{kim2022self,
  title={Self-serviced iot: Practical and private iot computation offloading with full user control},
  author={Kim, Dohyun and Patidar, Prasoon and Zhang, Han and Anilkumar, Abhijith and Agarwal, Yuvraj},
  journal={arXiv preprint arXiv:2205.04405},
  year={2022}
}

@article{tian2023chatgpt,
  title={Is ChatGPT the ultimate programming assistant--how far is it?},
  author={Tian, Haoye and Lu, Weiqi and Li, Tsz On and Tang, Xunzhu and Cheung, Shing-Chi and Klein, Jacques and Bissyand{\'e}, Tegawend{\'e} F},
  journal={arXiv preprint arXiv:2304.11938},
  year={2023}
}

@article{kokolakis2017privacy,
  title={Privacy attitudes and privacy behaviour: A review of current research on the privacy paradox phenomenon},
  author={Kokolakis, Spyros},
  journal={Computers \& security},
  volume={64},
  pages={122--134},
  year={2017},
  publisher={Elsevier}
}

@inproceedings{redmiles2019well,
  title={How well do my results generalize? comparing security and privacy survey results from mturk, web, and telephone samples},
  author={Redmiles, Elissa M and Kross, Sean and Mazurek, Michelle L},
  booktitle={2019 IEEE Symposium on Security and Privacy (SP)},
  pages={1326--1343},
  year={2019},
  organization={IEEE}
}

@misc{Permissi36:online,
author = {Google},
  title = {Permission Denials | App quality | Android Developers},
  url = "https://developer.android.com/topic/performance/vitals/permissions",
month = {04},
year = {2025},
  note = "[Online; accessed 2025-04-10]"
}

@inproceedings{khandelwal2021prisec,
  title={$\{$PriSEC$\}$: a privacy settings enforcement controller},
  author={Khandelwal, Rishabh and Linden, Thomas and Harkous, Hamza and Fawaz, Kassem},
  booktitle={30th USENIX Security Symposium (USENIX Security 21)},
  pages={465--482},
  year={2021}
}

@inproceedings{fernandes2018decentralized,
  title={Decentralized action integrity for trigger-action IoT platforms},
  author={Fernandes, Earlence and Rahmati, Amir and Jung, Jaeyeon and Prakash, Atul},
  booktitle={Proceedings 2018 Network and Distributed System Security Symposium},
  year={2018}
}

@inproceedings{cao2024stateful,
  title={Stateful least privilege authorization for the cloud},
  author={Cao, Leo and Meng, Luoxi and Stefan, Deian and Fernandes, Earlence},
  booktitle={33rd USENIX Security Symposium (USENIX Security 24)},
  pages={3477--3494},
  year={2024}
}

@inproceedings{chen2022practical,
  title={Practical data access minimization in $\{$Trigger-Action$\}$ platforms},
  author={Chen, Yunang and Alhanahnah, Mohannad and Sabelfeld, Andrei and Chatterjee, Rahul and Fernandes, Earlence},
  booktitle={31st USENIX Security Symposium (USENIX Security 22)},
  pages={2929--2945},
  year={2022}
}

@article{de2014openpds,
  title={openpds: Protecting the privacy of metadata through safeanswers},
  author={De Montjoye, Yves-Alexandre and Shmueli, Erez and Wang, Samuel S and Pentland, Alex Sandy},
  journal={PloS one},
  volume={9},
  number={7},
  pages={e98790},
  year={2014},
  publisher={Public Library of Science San Francisco, USA}
}

@inproceedings{zhang2021capture,
  title={Capture: Centralized library management for heterogeneous $\{$IoT$\}$ devices},
  author={Zhang, Han and Anilkumar, Abhijith and Fredrikson, Matt and Agarwal, Yuvraj},
  booktitle={30th USENIX Security Symposium (USENIX Security 21)},
  pages={4187--4204},
  year={2021}
}

@inproceedings{zhang2023don,
  title={Don't leak your keys: Understanding, measuring, and exploiting the appsecret leaks in mini-programs},
  author={Zhang, Yue and Yang, Yuqing and Lin, Zhiqiang},
  booktitle={Proceedings of the 2023 ACM SIGSAC Conference on Computer and Communications Security},
  pages={2411--2425},
  year={2023}
}

@article{shi2025skeleton,
  title={The Skeleton Keys: A Large Scale Analysis of Credential Leakage in Mini-apps},
  author={Shi, Yizhe and Yang, Guangliang and Yang, Zhemin and Yang, Yifan and Yang, Min and Zhong, Kangwei and Zhang, Xiaohan},
  year={2025}
}

@inproceedings{tahaei2023stuck,
  title={Stuck in the permissions with you: Developer \& end-user perspectives on app permissions \& their privacy ramifications},
  author={Tahaei, Mohammad and Abu-Salma, Ruba and Rashid, Awais},
  booktitle={Proceedings of the 2023 CHI Conference on Human Factors in Computing Systems},
  pages={1--24},
  year={2023}
}

@inproceedings{micinski2017user,
  title={User interactions and permission use on android},
  author={Micinski, Kristopher and Votipka, Daniel and Stevens, Rock and Kofinas, Nikolaos and Mazurek, Michelle L and Foster, Jeffrey S},
  booktitle={Proceedings of the 2017 CHI Conference on Human Factors in Computing Systems},
  pages={362--373},
  year={2017}
}

@inproceedings{alsoubai2022permission,
  title={Permission vs. app limiters: profiling smartphone users to understand differing strategies for mobile privacy management},
  author={Alsoubai, Ashwaq and Ghaiumy Anaraky, Reza and Li, Yao and Page, Xinru and Knijnenburg, Bart and Wisniewski, Pamela J},
  booktitle={Proceedings of the 2022 CHI Conference on Human Factors in Computing Systems},
  pages={1--18},
  year={2022}
}

@inproceedings{tahaei2022recruiting,
  title={Recruiting participants with programming skills: A comparison of four crowdsourcing platforms and a CS student mailing list},
  author={Tahaei, Mohammad and Vaniea, Kami},
  booktitle={Proceedings of the 2022 CHI Conference on Human Factors in Computing Systems},
  pages={1--15},
  year={2022}
}

@article{kroschewski2023save,
  title={Save the implicit flow? Enabling privacy-preserving RP authentication in OpenID connect},
  author={Kroschewski, Maximilian and Lehmann, Anja},
  journal={Proceedings on Privacy Enhancing Technologies},
  year={2023}
}

@article{corre2017can,
  title={Why can’t users choose their identity providers on the web?},
  author={Corre, Kevin and Barais, Olivier and Suny{\'e}, Gerson and Frey, Vincent and Crom, Jean-Michel},
  journal={Proceedings on Privacy Enhancing Technologies},
  volume={2017},
  number={3},
  pages={72--86},
  year={2017}
}

@article{zufferey2023revoked,
  title={“Revoked just now!” Users’ Behaviors toward Fitness-Data Sharing with Third-Party Applications},
  author={Zufferey, No{\'e} and Niksirat, Kavous Salehzadeh and Humbert, Mathias and Huguenin, K{\'e}vin},
  journal={Proceedings on Privacy Enhancing Technologies},
  volume={1},
  pages={47--67},
  year={2023}
}

@article{zavalyshyn2022sok,
  title={SoK: Privacy-enhancing Smart Home Hubs},
  author={Zavalyshyn, Igor and Legay, Axel and Rath, Annanda and Rivi{\`e}re, Etienne},
  journal={Proceedings on Privacy Enhancing Technologies},
  volume={4},
  pages={24--43},
  year={2022}
}

@misc{PiholeN58:online,
author = {Pi-hole},
  title = {Pi-hole – Network-wide Ad Blocking},
  url = "https://pi-hole.net/",
month = {11},
year = {2025},
  note = "[Online; accessed 2025-11-30]"
}

@inproceedings{jannett2024sok,
  title={Sok: Sso-monitor-the current state and future research directions in single sign-on security measurements},
  author={Jannett, Louis and Westers, Maximilian and Wich, Tobias and Mainka, Christian and Mayer, Andreas and Mladenov, Vladislav},
  booktitle={2024 IEEE 9th European Symposium on Security and Privacy (EuroS\&P)},
  pages={173--192},
  year={2024},
  organization={IEEE}
}

@misc{googleconfidential:Online,
	title = {Introducing Google Cloud Confidential Computing with Confidential VMs},
    author = {Nelly Porter and Sam Lugani},
	howpublished = {\url{https://cloud.google.com/blog/products/identity-security/introducing-google-cloud-confidential-computing-with-confidential-vms}},
	year = {2020},
    month = {07},
	note = {[Accessed 02-15-2026]},
}

@misc{vaultwarden:Online,
	title = {Unofficial Bitwarden compatible server written in Rust, formerly known as bitwarden\_rs},
    author = {Daniel García},
	howpublished = {\url{https://github.com/dani-garcia/vaultwarden}},
	year = {2022},
	note = {[Accessed 02-15-2026]},
}

\appendix%TC:ignore

\section{Analysis of Data Access in OAuth Apps}\label{sec:oauthapps}

\begin{table*}[t!]
    \centering\small
    \begin{tabular}{|p{23mm}| p{56mm} | p{46mm}| p{38mm}|}
    \hline
    \textbf{Example App} & \textbf{App Functionality} & \textbf{Current Privilege} & \textbf{Least Privilege}\\
    \hline\hline
    Uber Travel & Scan email for flights to book ride-share trips on arrival & All emails; Email settings (filters, labels) & Flight itinerary details\\
    \hline
    WellyBox & Scan email for financial receipts for reimbursement filing & All emails; Email settings (filters, labels) & Receipt attachments in emails \\
    \hline
    TripIt & Organize travel itinerary and information & All emails; Email settings (filters, labels) & Transportation and accommodation information in emails \\
    \hline
    Zoom & View upcoming video conferencing meetings; Create video conferencing meetings; See contacts on Zoom. & View, edit, and delete events on all calendars; See, edit, and delete all accessible calendars; See, edit, and permanently delete contacts & Title, time, and video conferencing links of events; Create events, not delete; See or add contacts, not modify \\
    \hline
    Doodle & Scheduling meetings with video conferencing and support for event invites on Google Calendar. & See and download any calendar you can access using your Google Calendar; View and edit events on all your calendars & Time availability blocks; Create events, not delete or edit \\
    \hline
    Notability & Note taking app with backups on Google Drive & See, edit, create, and delete all of your Google Drive files & See, edit, create, and delete Notability files on Google Drive \\
    \hline
    Slack Google Drive Add-on & Slack integration for Google Drive actions like file sharing and creation & See, edit, create, and delete all of your Google Drive files & Create files or modify file sharing permissions \\
    \hline
    Zotero GoogleDocs Integration & Edit citations in a specific file in Google Docs& See, edit, create, and delete all of your Google Drive files & Edit files; See the citations used in the files\\
    \hline
    Canvas & Attach Google Docs in assignment submissions & See, edit, create, and delete all of your Google Drive files & See files in Google Drive \\
    \hline
    Wordpress & Access to Google Photos to attach to posts & View your Google Photos library & View specific photos or album \\
    \hline
    Lenovo Smart Frame & Access to Google Photos to display on the frame & View your Google Photos library & View specific photos or album \\
    \hline
    \end{tabular}
    \caption{We analyzed 62 OAuth apps from Google Marketplace. Most apps request more privileges than they need to perform their tasks, and the data minimization needs are diverse and app-dependent. 
    }
    \label{tab:data-overaccess-cases}
    \vspace{-10pt}
\end{table*}

We analyzed 11 randomly selected OAuth apps using browser inspection and dynamic instrumentation to validate our assumptions about their data access (see Table~\ref{tab:data-overaccess-cases}).

% Because OAuth data transfers typically occur between back-end servers, client-side observations may not fully reflect actual data access. We assume the client first retrieves raw data from its back-end server and then processes it locally; however, any additional data requests made server-side are not visible from user devices. Despite this limitation, our analysis still provides empirical insights into OAuth data access in practice.

\sssec{Method}. We randomly selected 11 apps from the set analyzed in Section~\ref{sec:understand}  using a computer program and manually installed each of them. These apps include both open-source and closed-source applications. For every app, we created dedicated test accounts, completed the OAuth authorization process, and systematically interacted with the application's core features to trigger representative OAuth data access.

Because OAuth data transfers between an app's back-end server and Google's resource server are not observable from the client side, we employed three complementary analysis techniques based on each app's platform and source availability. 
For open-source applications (Zotero, Canvas, and WordPress), we conducted direct source code analysis, examining the server-side code that interfaces with Google's resource APIs to identify the exact endpoints invoked and the data fields accessed. For closed-source web-based applications, we used Chrome Developer Tools to inspect network traffic after OAuth authorization. We monitored fetch requests and corresponding responses to identify the invoked API endpoints, examine response payloads, and determine which user attributes were delivered to the client. For mobile applications, we instrumented the apps at runtime using Frida to intercept network API requests and relevant client-side data-processing routines, allowing us to determine which user attributes were accessed and how they were utilized locally. 

% our browser inspection and dynamic instrumentation results may underestimate actual data access. Source code analysis of open-source apps does not suffer from this limitation, as it reveals the complete set of API calls made from the app.

% \revise{
% We employed three complementary analysis techniques based on each app's platform and source availability. 

% % Across all techniques, we recorded the OAuth scopes each app requested during authorization and compared these against the data the app actually accessed.
% }

% \revise{We randomly selected 11 apps from the set analyzed in Section~\ref{sec:understand} and manually installed each of them. For every app, we created dedicated test accounts, completed the OAuth authorization process, and systematically interacted with the application’s core features to trigger representative OAuth data access. For web-based applications, we used Chrome Developer Tools to inspect network traffic after OAuth authorization. We monitored fetch requests and corresponding responses to identify the invoked API endpoints, examine response payloads, and determine which user attributes were delivered to the client. For mobile applications, we instrumented the apps at runtime using Frida tool to intercept network API requests and relevant client-side data-processing routines, allowing us to determine which user attributes were accessed and how they were utilized locally.}

\sssec{Results}. Our findings confirm our analysis results in Section~\ref{sec:understand}. 
% Across all investigated apps, we did not find any evidence contradicting our data-access assumptions. 
Source code analysis of three open-source applications directly confirms that actual API usage aligns with our least-privilege characterization. For example, Zotero's Google Docs integration asks for full CRUD access to all Drive files, even though it only invokes endpoints to read and edit citations within specific documents. Similarly, Canvas  requests full Google Drive permissions, but only performs read operations for file selection during assignment submission.

For closed-source apps, since we cannot access their server, we can only observe partial evidence of OAuth data access from the client side. This partial evidence also confirms our assumptions. In Doodle, the website retrieves calendar event objects via its API, whose structure closely mirrors raw Google Calendar API responses (including titles, descriptions, and timestamps), then processes them locally to display only time-availability blocks. This suggests that Google’s current OAuth scopes lack sufficient data granularity, requiring apps to fetch raw data and filter it on the client side.
For WellyBox, the client app receives pre-processed data from its backend containing only receipt-relevant information, indicating server-side filtering prior to transmission.

\section{Manifest in Our Case Study}
% \sssec{Uber accesses Gmail content}
% \begin{lstlisting}[style=Manifest, label=lst:notability]
% TITLE: Uber
% DESCRIPTION: Get only flight related dates
% PIPELINE: Gmail -> SelectMessage -> FilterFlights 
%           -> ExtractDate -> SendToUber
% Gmail(type: "Pull", resourceType: "gmail", 
%       query: "{message(userId){snippet}}")
% SelectMessage(type: "Select", field: "messages")
% FilterFlights(type: "Filter", operation: "include", 
%               field: ["snippet"], targetValue: "flight")
% ExtractDate(type: "Extract", operation: "match", 
%             field: ["snippet"] ,pattern: "Datetime")
% SendToUber(type: "Post", destination: "https:api.uber.com/v1/")
% \end{lstlisting}

\sssec{Zoom accesses Google Calendar}.
\begin{lstlisting}[style=Manifest,label=lst:zoom]
TITLE: Zoom
DESCRIPTION: Get all upcoming Zoom meetings
PIPELINE: PullCalendarEvents -> SelectEvents -> FilterTime 
                             -> FilterZoom -> PostToZoom

PullCalendarEvents(type: "Pull", resourceType: "google_calendar", 
                   query: "{ events(calendarId) {...EventDetails} }")
SelectEvents(type: "Select", field: "events")
FilterTime(type: "Filter", operation: ">", 
           field: "start.dateTime", targetValue: NOW)
FilterZoom(type: "Filter", operation: "match", 
           field: ["location", "description"], 
           pattern: "zoom\.us", requirement: "any")
PostToZoom(type: "Post", destination: "www.zoom.us")
\end{lstlisting}

\sssec{Notability write Google Drive}
\begin{lstlisting}[style=Manifest, label=lst:notability]
TITLE: Notability
DESCRIPTION: Backup notes to Google Drive
PIPELINE: ReceiveRequest -> FilterPath -> Upload

ReceiveRequest(type: "Receive", source: "www.notability.com")
FilterPath(type: "Filter", operation: "match", field: ["parents"], 
           targetValue: "folderId")
Upload(type: "Write", action: "create", resourceType: "google_drive")
\end{lstlisting}
\vspace{-2em}
\label{fig:notability}

\section{Selected Applications For Analysis}\label{sec:applist}
The applications below are used in the OAuth overaccess study. 
These applications were sampled from authors' Google's management interface and Google Workplace Marketplace's recommendation page, covering common scenarios of OAuth access to various Google services. 

\subsection{Google OAuth Apps}
\begin{enumerate*}
    \item Viber
    \item WhatsApp Messenger
    \item Figma
    \item Internet Archive
    \item Quora
    \item Typing.io
    \item penpot
    \item Uber Travel
    \item WellyBox
    \item TripIt
    \item Zoom
    \item Doodle
    \item Notability
    \item Slack Google Drive Add-on
    \item Zotero Google Docs Integration
    \item Canvas
    \item Wordpress
    \item Draw.io
    \item Hypatia Create
    \item Pokemon Go
    \item EasyBib Bibliography Creator
    \item Lumio
    \item Flubaroo
    \item Form Ranger
    \item Colaboratory
    \item Pear Deck
    \item CloudConvert
    \item MathType
    \item Lucidchart
    \item DocHub
    \item ZIP Extractor
    \item Autocrat
    \item SketchUp for Schools
    \item CoRubrics
    \item OrangeSlice
    \item Auto-LaTeX Equations
    \item Grade Reports
    \item Highlight Tool
    \item CLOZEit
    \item Easy Accents
    \item Slides Translator
    \item Automagical Forms
    \item Kaizena
    \item Lenovo Smart Frame
    \item Spark Email Client
    \item Form Mule
    \item Doctopus
    \item Quilgo
    \item Kami
    \item Nearpod
    \item formLimiter
    \item Fluency Tutor for Google
    \item ClassReporter
    \item Slides Randomizer
    \item Classroom Attendance Tracker
    \item Timer for Google Forms
    \item Copy Down
    \item AutoProctor
    \item Doc Appender
    \item Classroom Manager
    \item Form Notifications
    \item Form Presenter+Timer
\end{enumerate*}

\subsection{Trello OAuth Apps}
\begin{enumerate*}
    \item Analytics \& Reports by Screenful
    \item Card Priority Badge
    \item API Developer ID Helper
    \item Binotel
    \item Hipporello Apps
    \item ScaledbyScreenful
    \item Habit Tracker by UpgradeYourBrain
    \item BigPicture
    \item ShowAttachments
    \item SmartDeadlines
    \item WhatsApp Notifications
    \item RewriteText
    \item SumUp
    \item Notes \& Docs
    \item Statuses Workflow 
    \item TeamsHub
    \item Durations
    \item Google Sheets + Trello Two-Way Sync
    \item ConfluenceTrello
    \item Office File Viewer
    \item Wordefy
    \item Miro
    \item CustomFields
    \item Localize Borads
    \item 3DViewer
    \item Notion
    \item ExportsbyBlueCat
    \item BulkActions
    \item Contalist
    \item Hourly
    \item Timeline, Calendar, Time Tracking by Planyway
    \item OKRStudio
    \item Litmus
    \item OfficeHub
    \item Crmble
    \item Costello
    \item Jira + Trello 2-Way Sync
    \item Attachments Archive
    \item BoardExport
    \item Tables \& Spreadsheet
    \item PipeDrive
    \item join.me
    \item GitLab
    \item Countdown
    \item Amazing Fields
    \item TrelloTimeTracking
    \item FormsbyBlueCat
    \item EmailforTrello
    \item TimeFlowTracker
    \item Notejoy
    \item Onedrive
    \item Marker.io
\end{enumerate*}
    
\subsection{Slack OAuth Apps}
\begin{enumerate*}
    \item BambooHR
    \item Stack Overflow for Teams
    \item G2 for Slack
    \item Trello
    \item Hootsuite
    \item ZoomInfo
    \item HubSpot
    \item Miro
    \item Live Chat
    \item Qualtrics
    \item Zapier
    \item TheNewYorkTimes
    \item JiraCloud
    \item Workday
    \item Polly
    \item Freshdesk
    \item EddyTravels
    \item Evergreen
    \item Lattice
    \item AWSChatbot
    \item Bitbucket Cloud
    \item Box
    \item OneDrive and SharePoint
    \item GitLab
    \item Typeform
    \item Front
    \item Donut
    \item monday.com
    \item Asana
    \item Dropbox
    \item Drift
    \item ActiveCampaign Bot
    \item SurveySparrow
    \item ConfluenceCloud
    \item Airtable
    \item PivotalTracker
    \item Sentry
    \item Intercom
    \item Zoom 
    \item Notion
    \item ClickUp
    \item Frame
    \item Todoist
    \item GitHub
    \item Coffice
    \item PagerDuty
    \item Favro
    \item Harvest
    \item SimplePoll
    \item SalesforceforSlack
    \item Linear
    \item Crashlytics
    \item latexbot
\end{enumerate*}

\subsection{Github OAuth Apps}
\begin{enumerate*}
    \item Jenkins
    \item Slack
    \item Choreo.dev
    \item Swarmia
    \item DeepSource
    \item CodeReviewBot.AI
    \item Backup Github (Backrightup)
    \item giscus
    \item Pullflow
    \item Sourcery
    \item SonarCloud
    \item CR.GPT
    \item Ally
    \item Stacked Pull Requests
    \item codebeat
    \item GraphQL Hive
    \item Apollo Studio
    \item Report.Ci
    \item ReadME API
    \item Dpulls
    \item autofix.ci
    \item SyncDepot
    \item Honeycomb.io
    \item Test Check Publisher
    \item PeerGraph
    \item Google Cloud Build
    \item Bytesafe Integration
    \item Faable Deploy
    \item gitpod.io
    \item XRPL Bot
    \item DevHub
    \item Octobox
    \item GitGuardian
    \item Auto Docs
    \item AppSweep
    \item QuickFastChange
    \item Localazy
    \item Sadaiv CI
    \item Cloudback: GitHub Backup \& Restore
    \item Back Git Up
    \item Telebirr-cloud-bot
    \item webapp.io
    \item harpoon.io
    \item Devbox Cloud
    \item Depfu
    \item Phylum
    \item PDDON
    \item GitBook
    \item StarChart Labs | Chronicler
    \item Payvost
    \item rootly
\end{enumerate*}

\section{Annotated Code Changes for \sysname Integration}
\label{sec:code_changes}
\begin{lstlisting}[style=diffstyle, label={lst:zoom-imports-diff}]
// src/pages/index.tsx
  ......
  // Include the manifest pipeline
+ const manifest = "<Zoom Manifest Text>`;

  const getAuthUrl = async () => {
-   const res = await fetch("/api/auth");
-   const { authUri } = await res.json() as { authUri?: string };
    
    // Authorization
+   const OAuthHubClient = require("~/lib/oauthub-client");
+   return OAuthHubClient.generateAuthUrl({
+       provider: "google_calendar", 
+       manifest: manifest,
+       redirect: "zoom.us/callback", 
+       accessType: "user_driven",
+   }) as Promise<string>;
  };

  const requestData = async () => {
-   await fetch("/api/data", {method:"POST"});
+   const OAuthHubClient = require("oauthub-client");
+   const authRes = await fetch("/api/oauthub/auth");
+   const { token } = await authRes.json() as { token: string | null };
+   const result = await OAuthHubClient.query({ token, manifest: manifest });
+   await fetch("/api/oauthub/auth", {
+     method: "POST",
+     headers: { "Content-Type": "application/json" },
+     body: JSON.stringify({token: result.token }),
+   });
  };
  ......
  
  // UI Changes
+ <OAuthHubIcon />
+ Sign in with OAuthHub
  ......
\end{lstlisting}

\begin{lstlisting}[style=diffstyle, label={lst:zoom-imports-diff}]
// src/pages/api/callback.tsx
   ......
   // Extract hub identity params
+  let publicKey = JSON.parse(params.get("oauthub_public_key");

   // Exchange auth code for access token 
+  const OAuthHubClient = require("~/lib/oauthub-client");
+  const { access_token } = await OAuthHubClient.exchangeToken({ code, state });

-  await fetch("/api/auth", {
-    body: JSON.stringify({ code }),

+  await fetch("/api/oauthub/auth", {
+    method: "POST",
+    headers: { "Content-Type": "application/json" }, 
+    body: JSON.stringify({ publicKey, userSub, token: access_token }),
   });
  ......
\end{lstlisting}

% \begin{lstlisting}[style=diffstyle, label={lst:zoom-imports-diff}]
% // src/pages/api/auth.ts
%    export default async function handler(req: NextApiRequest, res: NextApiResponse) {
% -    const { code } = req.body as { code: string };
% -    const { tokens } = await oauth2Client.getToken(code);
% -    await db.push("/google_tokens", { ...tokens });
% +    const { publicKey, userSub, token } = req.body;
% +    const existing = await db.getData("/oauthub");
% +    const update: Record<string, unknown> = { ...existing };
% +    await db.push("/oauthub", update);
%   }
% \end{lstlisting}

\begin{lstlisting}[style=diffstyle, label={lst:zoom-imports-diff}]
// src/pages/api/data.ts
  ......
+ const OAuthHubClient = require("oauthub-client");

  export default async function handler(req, res) {
-   const tokens = await db.getData("/google_tokens");
-   oauth2Client.setCredentials(tokens);
-   const calendar = google.calendar({ version: "v3", auth: oauth2Client});
-   const { data } = await calendar.events.list({
-     calendarId: "primary",
-     timeMin: new Date().toISOString(),
-     maxResults: 10,
-     singleEvents: true,
-     orderBy: "startTime",
-   });
-   const events = (data.items.map((event) => ({
-     summary: event.summary",
-     start: event.start.dateTime,
-        end: event.end.dateTime,
-   }));

+   const signature = req.headers["x-oauthub-signature"];
+   const body = req.body;
+   const rawBody = JSON.stringify(body);

    // Verify payload signature
+   let publicKeyJwk = await db.getData("/oauthub").publicKey;
+   if (signature && publicKeyJwk) {
+      const valid = await OAuthHubClient.verifyJWT({ jwt: signature, publicKeyJwk, body: rawBody });
+      if (!valid) return res.status(401).json({ error: "Invalid signature" });
+   }
+   const events = body.data;
  ......
\end{lstlisting}

\section{Survey Questions for the User Study}
\label{sec:survey}

\subsection{Zoom}
Zoom is a video conferencing app that allows users to connect for meetings, webinars, and virtual events. It supports high-quality video and audio, screen sharing, and collaboration tools like chat and file sharing. Zoom is widely used for remote work, education, and social interactions, providing an easy-to-use platform for connecting with others online.

Zoom can be integrated with Google Calendar so that users can schedule and join Zoom meetings directly from the calendar.

\vspace*{0.05in}\noindent [Displayed one of the following authorization interface conditions]

\sssec{Condition A}. When linking with Google Calendar, you'll be presented with an authorization page as shown below.

\begin{figure}[h!]
    \centering
    \includegraphics[width=0.6\linewidth]{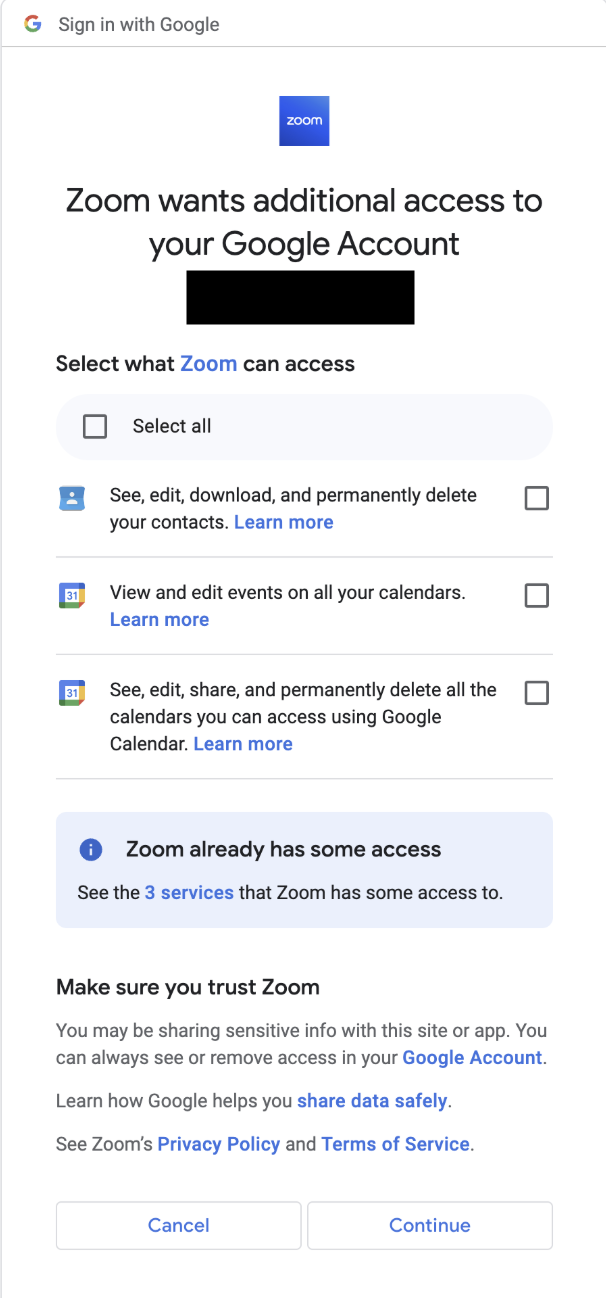}
    \vspace{-10pt}
\end{figure}

\sssec{Condition B}. When linking with Google Calendar through \sysname, you'll see an interactive authorization page below. If you grant the permission, \sysname will processes your personal data locally on your device and sends only the processed results to the app. The ``Overview'' provides a general illustration of the process. ``Filters'' gives a detailed step-by-step description. ``Raw data'' shows the data before and after processing.

\begin{figure}[h!]
    \centering
    \includegraphics[width=1.02\linewidth]{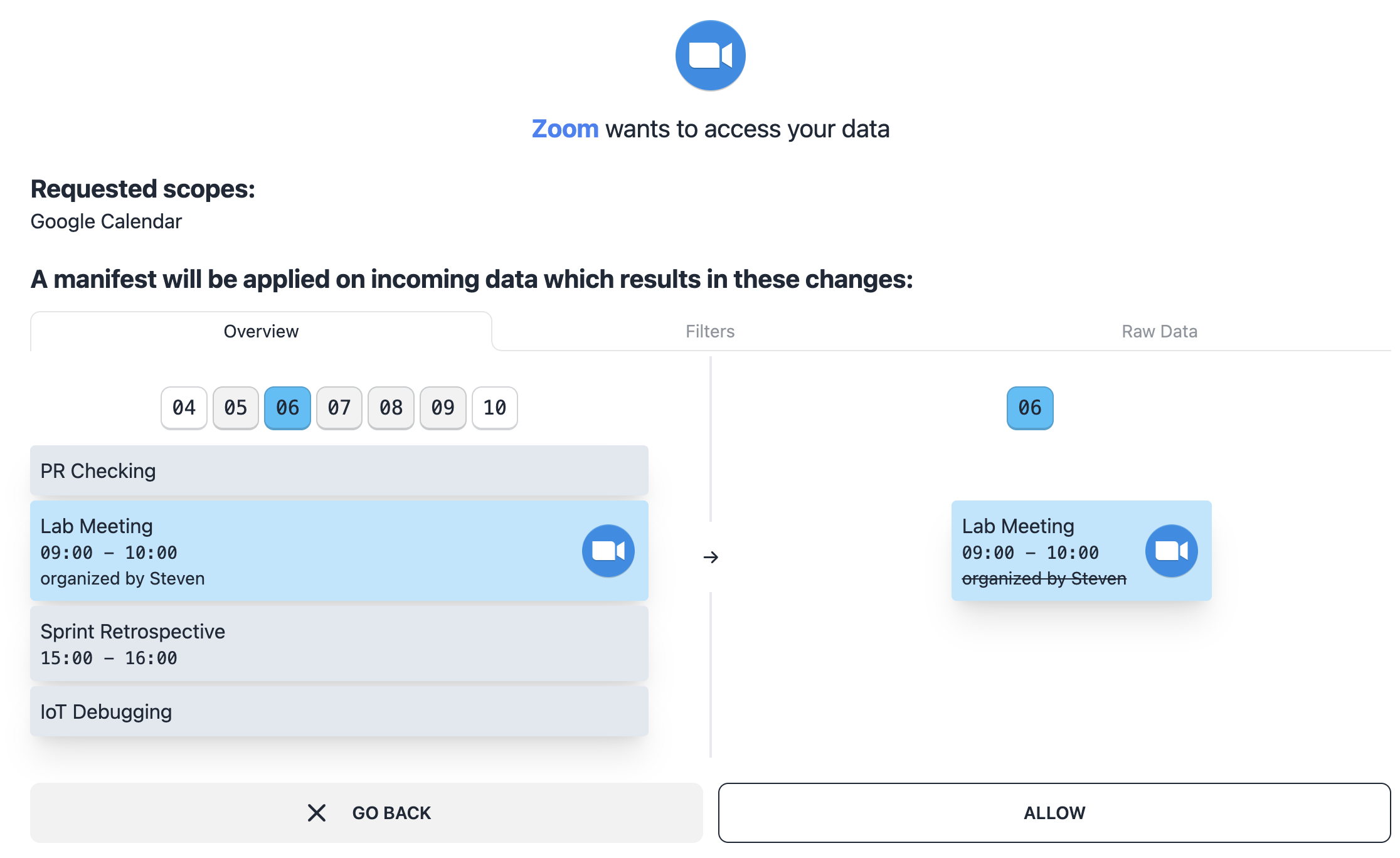}
    \vspace{-20pt}
\end{figure}

\vspace*{0.05in}\noindent [Present the following questions after the interface exposure]

\begin{enumerate}[1., leftmargin=*]
    \item If you grant Zoom the requested permission, what data will Zoom be able to access?
\begin{itemize}
    \item Events you explicitly share with others
    \item All of your calendar events
    \item All Zoom meeting events
    \item Zoom meeting topics, time, and links
\end{itemize}
    \item How likely would you grant the app the requested permission? (Choose from a 5-point Likert scale)
    \begin{itemize}
    \item Extremely unlikely
    \item Somewhat unlikely
    \item Neither likely nor unlikely
    \item Somewhat likely
    \item Extremely likely
    \end{itemize}
    \item Please share your thoughts on why you'd choose to grant or not grant Zoom these permissions. Do you have any unaddressed concerns? (Open-ended question)
\end{enumerate}

\subsection{Uber}
Uber is a ride-hailing platform that connects passengers with drivers for transportation services. Uber links with Gmail to automatically import and organize your travel plans, such as flight, hotel, and restaurant reservations, directly into the Uber app. By accessing your Gmail account, Uber can scan for travel-related emails and create a unified itinerary within the app.

\vspace*{0.05in}\noindent [Displayed one of the following authorization interface conditions]

\sssec{Condition A}. When linking with Gmail, you'll be presented with an authorization interface as shown below.

\begin{figure}[htbp]
    \centering
    \vspace{-20pt}
    \includegraphics[width=0.6\linewidth]{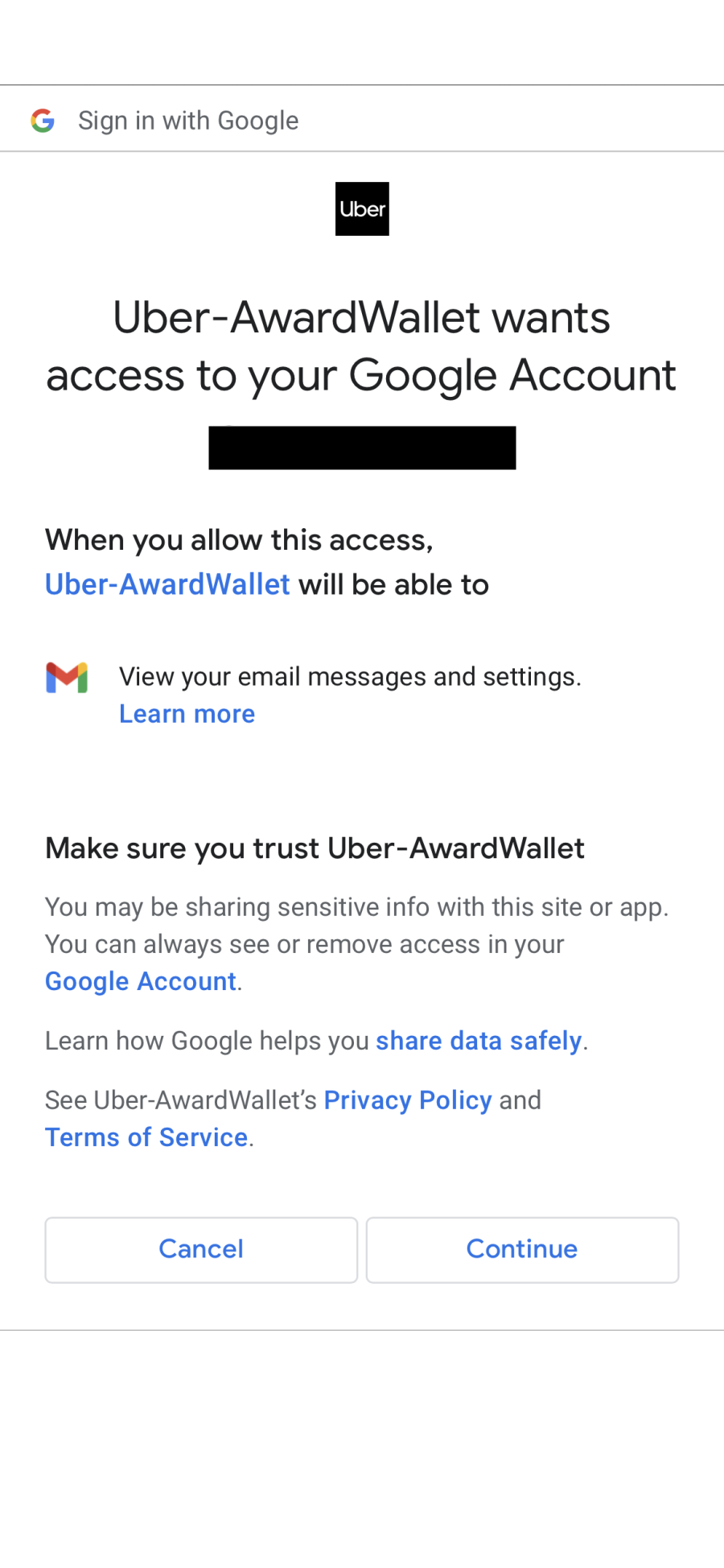}
    \vspace{-60pt}
\end{figure}

\sssec{Condition B}. When linking with Google Drive through \sysname, you'll see an interactive authorization page below. If you grant the permission, \sysname will process your personal data locally on your device and send only the processed results to the app. The ``Overview'' provides a general illustration of the process. ``Filters'' gives a detailed step-by-step description. ``Raw data'' shows the processed data.

\begin{figure}[h!]
    \centering
    \includegraphics[width=1.02\linewidth]{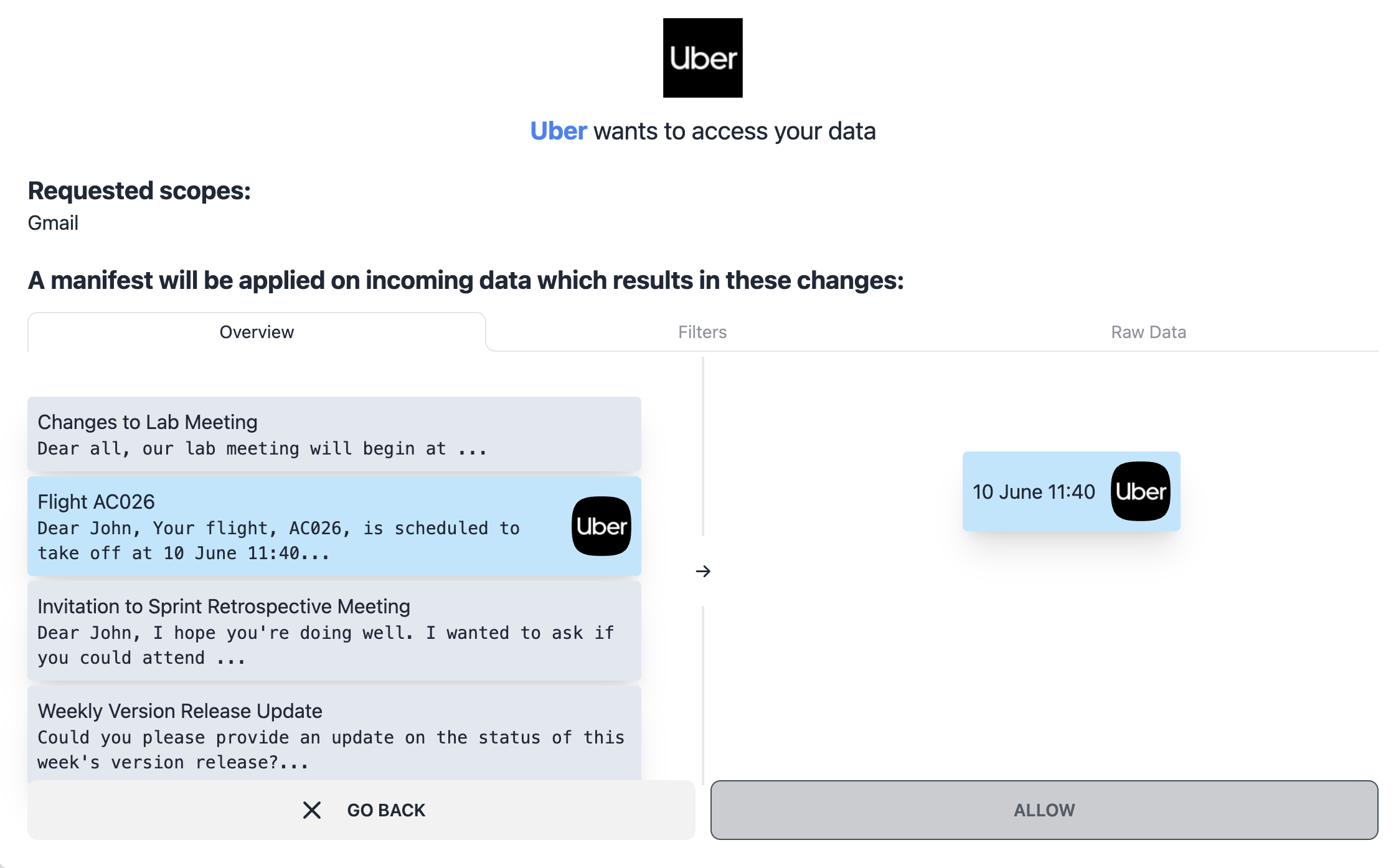}
\end{figure}

\vspace*{0.05in}\noindent [Present the following questions after the interface exposure]

\begin{enumerate}[1., leftmargin=*]
    \item If you grant Uber the requested permission, what data will Uber be able to access?
    \begin{itemize}
    \item Only emails with travel information
    \item All of your emails
    \item Only flight information extracted from emails
    \item All emails received from Uber
    \end{itemize}
    \item How likely would you grant the app the requested permission? (Choose from a 5-point Likert scale)
    \begin{itemize}
    \item Extremely unlikely
    \item Somewhat unlikely
    \item Neither likely nor unlikely
    \item Somewhat likely
    \item Extremely likely
    \end{itemize}
    \item Why would you choose to grant or not grant these permissions to Uber? Do you have any unaddressed concerns? (Open-ended question)
\end{enumerate}

\subsection{Notability}
Notability is a note-taking app designed for iOS and macOS that combines handwriting, typing, and audio recording into a single platform. The app supports multimedia, allowing users to add images, web clips, and more to their notes. Notability links with Google Drive to provide users with additional storage and backup options for their notes.

\vspace*{0.05in}\noindent [Displayed one of the following authorization interface conditions]

\sssec{Condition A}. When linking with Google Drive, you'll be presented with an authorization page as shown below.

\begin{figure}[h!]
    \centering
    \includegraphics[width=0.6\linewidth]{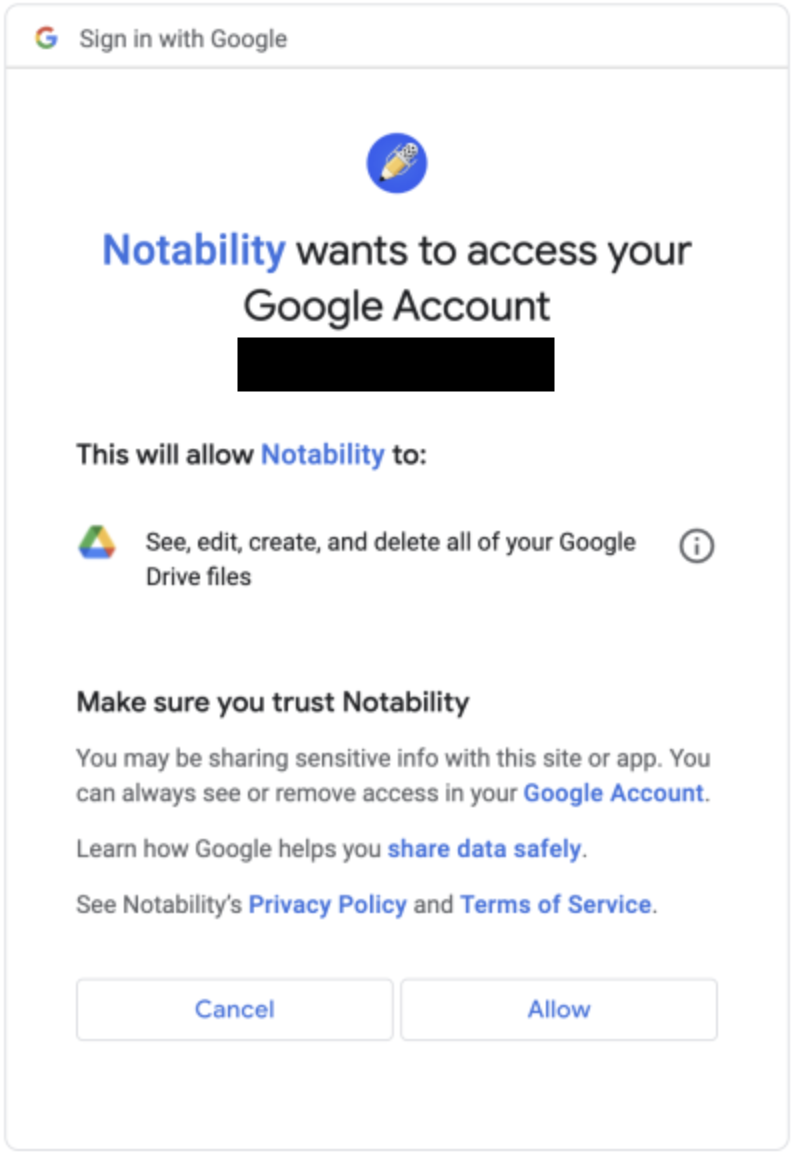}
    \vspace{-10pt}
\end{figure}

\sssec{Condition B}. When linking with Google Drive through \sysname, you'll see an interactive authorization page below. If you grant the permission, \sysname will processes your personal data locally on your device and sends only the processed results to the app. The ``Overview'' provides a general illustration of the process. ``Filters'' gives a detailed step-by-step description. ``Raw data'' shows the processed data.

\begin{figure}[h!]
    \centering
    \includegraphics[width=1.02\linewidth]{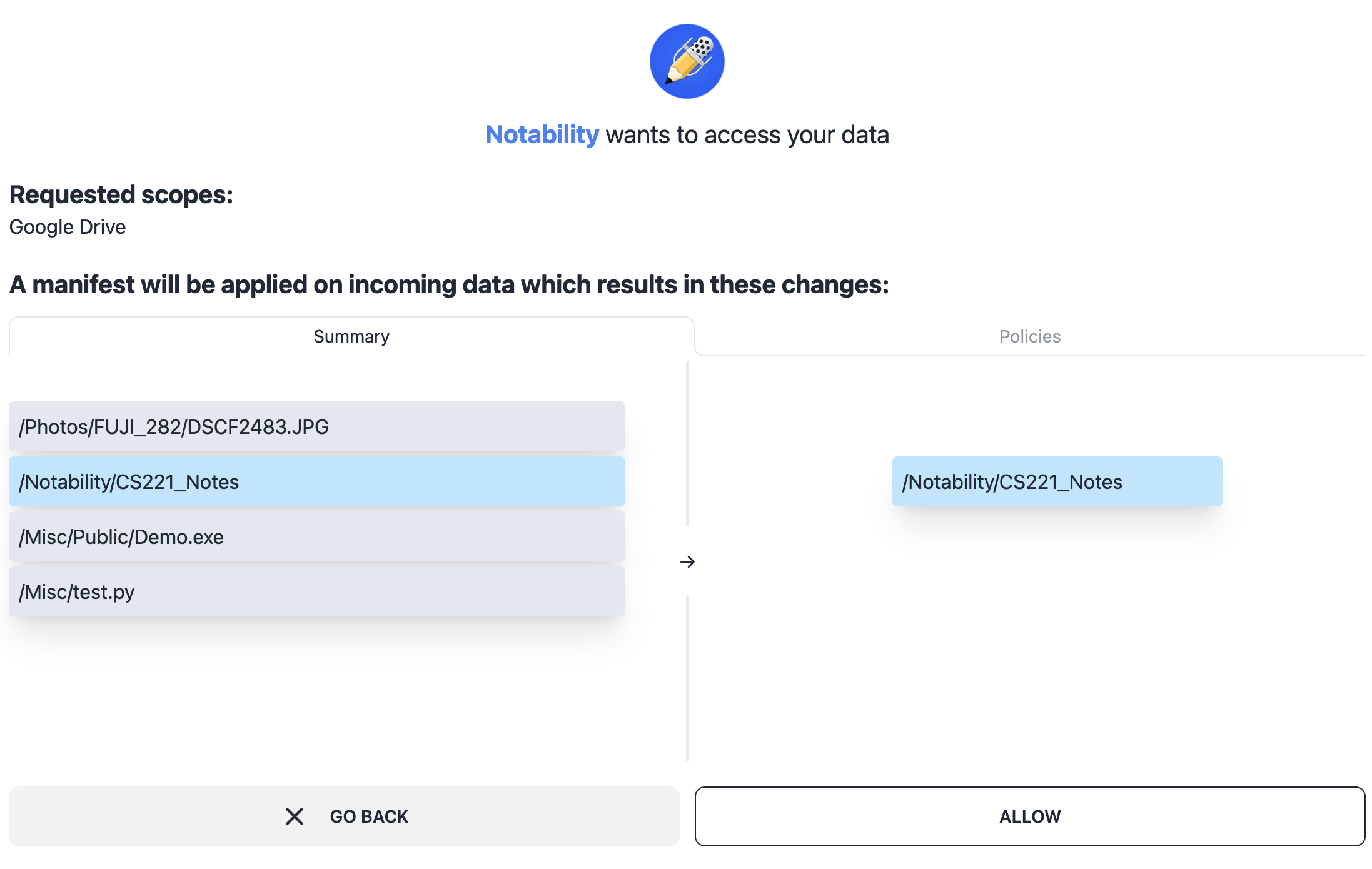}
\end{figure}

\vspace*{0.05in}\noindent [Present the following questions after the interface exposure]

\begin{enumerate}[1., leftmargin=*]
    \item If you grant Notability the requested permission, what data will Notability be able to access?
\begin{itemize}
    \item Files created by Notability
    \item All Images and documents
    \item Files under the Notability folder
    \item All Google Drive files
\end{itemize}
    \item How likely would you grant the app the requested permission? (Choose from a 5-point Likert scale)
    \begin{itemize}
    \item Extremely unlikely
    \item Somewhat unlikely
    \item Neither likely nor unlikely
    \item Somewhat likely
    \item Extremely likely
    \end{itemize}
    \item Why would you choose to grant or not grant these permissions to Notability? Do you have any unaddressed concerns? (Open-ended question)
\end{enumerate}

\begin{table*}[t]
    \centering\small
    \begin{tabular}{|p{12mm}| p{12mm}| p{48mm}| p{65mm}|}
    \hline
    Category & Operator & Parameters & Description \\
    \hline    \hline
    \multirow{2}{*}{Provider} & Pull & resourceType & Get data from an OAuth resource\\\cline{2-4} 
     & Receive & source & Accept action requests from OAuth apps\\
    \hline
    \multirow{4}{*}{Reduction} & Filter & operation, requirement, field, targetValue & Filter data based on specific conditions on certain field(s) \\\cline{2-4}
    & Select & field & Select specific field(s) in the data \\\cline{2-4} 
    & Extract &  operation, field, pattern & Extract data in a defined pattern from a certain field\\\cline{2-4} 
    & Limit & count & Limit the number of results returned\\
    % & Restrict & path, fileTypes & Works with the Write operator to restrict write path and file types\\\cline{2-4} 
    \hline
    \multirow{3}{*}{Transform} & Aggregate & operation, field, groupKey & Compute aggregate data for specific field \\\cline{2-4} 
    & Map & operation, field & Apply a mapping operation to a specific field \\\cline{2-4} 
    % & Noisify &  dataType, target, likertMean, likertStdDev, rangeType & Inject noise to the data\\\cline{2-4} 
    % & Sort &  sortKey, order & Sort results according to a particular order\\\cline{2-4} 
    & Anonymize &  dataType, field, method & Anonymize data in a specific field using a defined method \\
    \hline
    \multirow{2}{*}{Network} & Post & destination & Send the processed results to OAuth apps \\\cline{2-4} 
    & Write & action, resourceType & Perform actions on a specified OAuth resource\\
    \hline
    \multirow{2}{*}{Utility} & Inject & repeatNum, interval & Trigger the execution periodically \\
    \cline{2-4}
    & Debug & & Print output data for debugging \\ 
    % Replace certain field with an artificial replacement value
    \hline
    \end{tabular}
    \caption{The taxonomy of \sysname operators. Each operator corresponds to a ``verb'' statement relative to its semantics.}
    \label{tab:oauthwall-operators}
\end{table*}

\subsection{Demographics and Background}
\begin{enumerate}[1., leftmargin=*]
    \item What is your gender?
\begin{itemize}
    \item Male
    \item Female
    \item Non-binary
    \item Prefer to self-describe: \underline{\qquad}
    \item Prefer not to answer
\end{itemize}
    \item What is your age?
    \begin{itemize}
        \item 18--24
        \item 25--34
        \item 35--44
        \item 45--54
        \item 55--64
        \item 65+
    \end{itemize}
    \item What is the highest level of education you have completed?
    \begin{itemize}
        \item High school or below
        \item Some college
        \item Associate degree
        \item Bachelor degree
        \item Master degree
        \item Doctoral degree
        \item Prefer not to answer
    \end{itemize}
    \item Please select the statement that best describes your comfort level with computing technology.
    \begin{itemize}
        \item I can build my own computers, run my own servers, code my own apps, etc.
        \item I know my way around computers and mobile/IoT devices pretty well; I am the person who helps friends and family with technical problems.
        \item I know how to use computers and mobile/IoT devices to perform my job and life responsibilities; I often need technical help from others.
        \item Technology usually scares me. I only use it when I have to.
    \end{itemize}
    \item Indicate the extent to which you agree or disagree with the following statements. 
    \textit{[For each statement below, participants were asked to choose from a 5-point likert scale ``Strongly disagree'' ``Somewhat disagree'' ``Neither agree or disagree'' ``Somewhat agree'' or ``Strongly agree.'' Statements were presented in randomized order.]}
    \begin{itemize}
        \item In general, I believe privacy is important
        \item In today’s world, privacy does not exist anymore.
        \item I’m concerned that “smart” devices collect too much personal information.
        \item I’m confident in my ability to control how much personal information I share online and through my phone.
    \end{itemize}
\end{enumerate}

\section{Example OAuth Screenshots}
\begin{figure}[htbp]
  \includegraphics[width=0.47\textwidth]{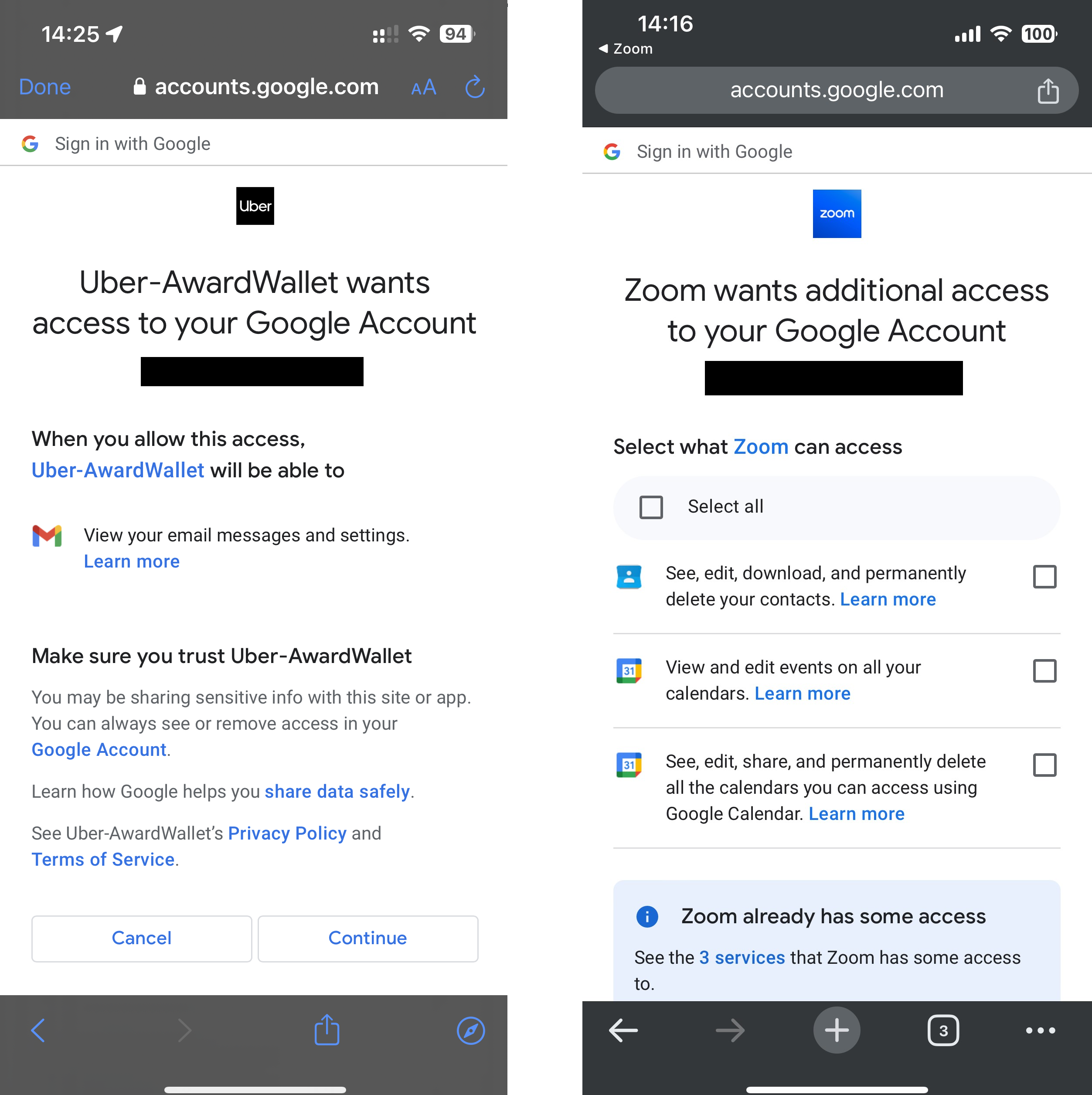}
  \caption{OAuth applications often request more data access than they need. Uber AwardWallet requires access to all users' emails, although it only needs to access flight confirmation emails to extract users' itineraries (Left). Zoom requests both read and write access to users' calendars and contacts (Right).
  }
  \label{fig:oauth-exampls}
\end{figure}

See Figure~\ref{fig:oauth-exampls}.

\label{sec:screenshots}

\section{Operator Design}
\label{sec:operators}
See Table~\ref{tab:oauthwall-operators}.

\appendix

\end{document}